\documentclass[letterpaper, 10 pt, conference]{ieeeconf}  

\IEEEoverridecommandlockouts                              
               \usepackage{array}
   \usepackage[pdftex]{graphicx}
   \usepackage{epstopdf}
   \usepackage{ float}
\usepackage[cmex10]{amsmath}
\usepackage{amsfonts}
\usepackage{times,amssymb, url,color,,soul }
\usepackage{comment,multirow}
\DeclareMathOperator*{\argmax}{\arg\!\max}

\usepackage{adjustbox}
\usepackage{rotating}
\newenvironment{boxedcode}
 {\setlength{\fboxsep}{.5em}%
   \begin{adjustbox}{minipage=\textwidth-2\fboxsep-2\fboxrule,fbox}
  \begin{tabbing}}
 {\end{tabbing}
  \end{adjustbox}}                                           

\title{\LARGE \bf
 Affective Movement Generation using Laban Effort and Shape and Hidden Markov Models}

\author{Ali-Akbar Samadani, Rob Gorbet, Dana Kuli\'c
\thanks{A. Samadani and  D. Kuli\'{c} are with the Department of Electrical and Computer Engineering, University of Waterloo, Waterloo, Canada.\newline
E-mail: asamadan@uwaterloo.ca,
 dkulic@uwaterloo.ca.}
\thanks{R. Gorbet is with the Department of Knowledge Integration, University of Waterloo, Waterloo, Canada. E-mail: rbgorbet@uwaterloo.ca}}

\begin{document}

\maketitle
\thispagestyle{empty}
\pagestyle{empty}

\begin{abstract}

Body movements are an important communication medium through which affective states can be discerned. Movements that convey affect can also give machines life-like attributes and help to create a more engaging human-machine interaction. This paper presents an approach for automatic affective movement generation that makes use of two movement abstractions:  1) 
 Laban movement analysis (LMA), and 2) hidden Markov modeling. The LMA  provides a systematic tool for an abstract representation of the kinematic and expressive characteristics of movements. 
Given a desired motion path on which a target emotion is to be overlaid, the proposed approach searches a labeled dataset in the LMA Effort and Shape space for 
similar movements to the desired motion path that convey the target emotion. An HMM abstraction of the identified movements is obtained and used with the desired motion path 
to generate a novel movement  that is a modulated version of the desired motion path that conveys the target emotion. The extent of modulation can be varied, trading-off between kinematic and affective constraints in the generated movement. The proposed approach is tested using a full-body movement dataset.
The efficacy of the proposed approach in generating movements with recognizable target emotions is assessed using a validated automatic recognition model and a user study. The target emotions were correctly recognized from the generated movements at a rate of 72\% using the recognition model. Furthermore, participants in the user study were able to correctly perceive the target emotions from a sample of generated movements, although some cases of confusion were also observed.

\end{abstract}
\section{Introduction}
\label{sec:intro}
 The ability to express affect via movements has the potential to improve the perceived intelligence and empathic attributes of autonomous agents (e.g., \cite{breazeal2004designing, Pineau2003271, Hartmann2006}) 
 and create a more engaging human-machine interaction. 
Affective movements are commonly generated by designers and animators using techniques such as keyframing \cite{Lasseter:1987}. These techniques, however, are time consuming and require trained animators and designers. 
Computational models  that can automatically generate affective movements would enable the generation of automatic affective responses and allow machines to engage in a long-term empathic interaction with their users.  

Automated generation techniques must be able to handle the large variability in affective movement. On the one hand, an emotion can be communicated through a number of kinematically different movements, and on the other hand, there exist kinematically similar movements that convey  distinct emotions. Furthermore, there are interpersonal and stochastic differences in affective movement expressions. An automatic generation model should suppress the irrelevant sources of variation and exploit movement features salient to affective expression in adapting a desired motion path to convey a target emotion. In addition, the model should be capable of generating affective movements in a computationally efficient manner. This paper presents a new affective movement generation approach that aims to fulfil these requirements. 
The proposed approach derives two low-dimensional movement representations using the Effort and Shape components of Laban movement analysis (LMA) \cite{laban1947effort, laban1971mastery} and hidden Markov modeling, to isolate salient kinematic and affective characteristics of the movements based on which a desired affective movement is generated. 

The LMA   is a prominent  movement analysis and notation system that was developed for writing and analyzing dance choreography \cite{laban1947effort}.
In a previous work, an approach for quantifying LMA Effort and Shape components for hand and arm movements was presented and validated in comparison with annotations provided by a certified motion analyst \cite{SamadaniACII2013}. In this paper, the LMA quantification from \cite{SamadaniACII2013} is adapted for annotating full-body movements and extended to include additional LMA Shape descriptors. The LMA quantification is computationally efficient and allows for an efficient search for movements kinematically and/or affectively 
similar to a desired motion path. The efficacy of Effort and Shape components for motion indexing and retrieval in large and high-dimensional movement datasets has also been previously demonstrated in~\cite{motion-retrieval-i3d}.

The second movement abstraction used in this work is based on hidden Markov models (HMM)s. An HMM is robust to temporal variations (length and phase variations in movements), can handle the noise inherent in human movements, provides a stochastic model for the dynamics of motion trajectories, and is efficient to learn. 

Given a labeled movement dataset, to adapt a desired motion path to convey a target emotion, the proposed approach proceeds as follows: 1) using the LMA Effort and Shape abstraction\footnote{A preliminary version of the LMA Effort and Shape quantification approach has been reported in~\cite{SamadaniACII2013}.  This paper uses an adapted version of the quantification approach reported in~\cite{SamadaniACII2013} to formulate a motion generation approach.  The novel motion generation approach and its experimental validation are reported for the first time in this paper.}, nearest neighbours to the desired motion path that belong to the target emotion class are identified, 2) an HMM for the identified nearest neighbours along with the desired motion path is learned, 3) the most likely state sequence of the learned HMM for the desired motion path is obtained using the Viterbi algorithm, 4) a deterministic generation approach \cite{Kulić01072008} is used to generate a modulated version of the desired motion path that encodes the target emotion.

The proposed approach is tested with a full-body affective movement dataset. 
The expressivity of the generated movements is assessed objectively using an automatic recognition model \cite{SamadaniTHMS2014} and subjectively, in a user study where participants report their perception of the generated movements.  

This paper is organized as follows: Section \ref{sec:rel} reviews the  related work on the automatic generation of affective movements followed by the description of the proposed approach in Section \ref{sec:method}. Section \ref{sec:exp} describes the experimental setup. Experimental results are reported in Section \ref{sec:res}. The results and future directions for the present study are discussed in Section \ref{sec:dis} followed by Section~\ref{sec:limit} that discusses the limitations of the proposed approach. We close the paper with conclusions 
in Section \ref{sec:con}.

\section{Related Work}
\label{sec:rel}
In general, there are two main approaches for automatic affective movement generation: 1) rule-based, and 2) example-based. 
Rule-based approaches propose a set of motion generation/modification rules that are mainly driven by an intuitive mapping between movement features and different affective expressions (or motion styles).  On the other hand, example-based approaches generate a desired affective movement by concatenating or blending movement exemplars from a labeled dataset of movements. 

The success of rule-based approaches depends on a-priori knowledge of postural and motion cues salient to different affective expressions and how  these cues are incorporated into the rules in terms of quantitative variables that can be tuned to generate a desired affective movement. 
A drawback of the rule-based approaches is that they are designed for a specific  kinematic structure and a set of movements, and often are not readily adaptable to other structures and movements. While this limitation also exists for the example-based approaches, they do not rely on any deterministic rules and are capable of generating novel movement trajectories provided that adequate movement exemplars are available. Therefore, the success of the example-based approaches depends on the richness of the labeled  dataset.

Several examples of rule-based approaches can be found in the literature. A  mapping from LMA  Effort and Shape to upper-body movement kinematics is proposed in \cite{Chi:2000}. Correlations between Laban components and discrete emotions  are used to devise a mapping from discrete emotions to joint angles of a humanoid robot \cite{Masuda2010}. In another study, a set of motion modification rules adjust the velocity, the range of motion, and the basic posture for a given movement based on specified levels of the arousal and valence affective dimensions \cite{Nakagawa2009}. A set of expressivity parameters  derived from the psychology literature (overall activation, spatial extent, temporal extent, fluidity, power, repetition) are used to develop rules that select a gesture type and modify it to adopt a desired style (e.g., abrupt) for an embodied conversational agent \cite{Hartmann2006}.

Example-based approaches range from concatenation of labeled motion segments (e.g., \cite{Arikan:2002}) to sophisticated generative models learned from exemplar movements (e.g., \cite{TaylorNIPS2006}).  Human movements vary in length and are intrinsically high-dimensional, and learning in high-dimensional motion space suffers from the limitations associated with the ``curse of dimensionality''.
 Different approaches are reported that make use of a low-dimensional  representation of affective movements along with an inverse mapping to generate new human movements. 
For instance, movements can be  represented as a weighted linear combination of a set of basis functions (functional representation) \cite{unuma1995fourier, SamadaniSocial2013}. The human gait is generated by interpolating or extrapolating Fourier expansions of actual human movements corresponding to different locomotion types (walk and run) subject to kinematic constraints (step size, speed, and hip position) \cite{unuma1995fourier}. Then, a gait is imbued with a target emotion via superposition of the Fourier expansion of the gait with a Fourier characteristic function for the target affective gait. 
In another study, a functional representation of a set of affective hand movements is obtained by transforming the movements into \textit{B}-spline basis space \cite{SamadaniSocial2013}. Then, functional principal component analysis is used to obtain a lower-dimensional subspace that serves as the basis space for affective movement generation. The independent component analysis is used to extract styles from one motion and subsequently infuse an extracted style to another motion sequences to generate desired stylistic walking patterns~\cite{Shapiro:2006}.
A limitation of the basis function expansion approaches is that they require movements to be of equal length with aligned landmarks,  which is a difficult constraint to attain under natural conditions.  Furthermore, a user's input is required to select the appropriate components representing the desired style~\cite{SamadaniSocial2013, Shapiro:2006}.

Other example-based approaches learn a mapping between interpolation parameters and the human joint space to generate different motion trajectories. 
Mukai and Kuriyama \cite{Mukai:2005} employ a Gaussian process regression approach to learn such a mapping. For each action (e.g., arm reaching), a set of control parameters are manually selected (e.g., turning angle and height of the target for the reaching movement) and a mapping between the parameter space and high-dimensional joint space is learnt for that action. The resulting action-specific mapping is used to generate different styles of the action (e.g., reaching to different target positions).
Movement styles were annotated in terms of Laban Weight, Time, and Flow Effort for the movements available in a labeled dataset 
 \cite{Torresani2006NIPS}, and for each pair of kinematically similar movements, a new movement is generated 
  via space-time interpolation. Subsequently, a mapping between resulting motion styles and the interpolation parameters is learnt. For a test movement and a target style, a pair of movements in the labeled dataset are selected, whose blend is kinematically similar to the test movement and provides the best approximation of the target style \cite{Torresani2006NIPS}. Stylistic gait patterns were automatically generated using a hidden semi-Markov model whose state output and duration distributions are tuned based on a desired pair of speed and stride length using a multiple regression~\cite{YamazakiHSMM}.  In another study, stylistic walking patterns are generated via interpolation and extrapolation between between style-specific HMM parameters~\cite{Tilmanne2012}. In~\cite{Hsu:2005},   stylistic motions are generated using spatiotemporal warping of joint-specific motion styles extracted from pairs of stylistically-different motions represented by linear dynamical  models.   The studies \cite{Mukai:2005,Torresani2006NIPS,YamazakiHSMM, Tilmanne2012, Hsu:2005} do not explicitly address affective movement generation and are specifically designed for a single movement  type (e.g., gait).  
   
Another class of example-based approaches derives a stochastic generative modeling of the movements. Among stochastic generative approaches, those that derive a  mapping from a lower-dimensional hidden space to the movement space are popular in human motion analysis. Style machines, a parametric hidden Markov modeling approach, provide a stochastic modeling of movements parametrized by a style variable to be used to generate movements with a desired style \cite{Brand:2000}.  Style machines were tested with a set of similar movements  performed in different styles (biped locomotion, and choreographed ballet dance), and were not explicitly tested to generate affective movements \cite{Brand:2000}. 
A Gaussian process (GP) dynamical model was used to learn motion styles as low-dimensional continuous latent variables and their mapping to movement observations. Using the learned model, a new movement is generated by adjusting the latent variable to adopt a desired style \cite{Wang2008}. GP-based models are often limited to a small number of aligned and similar movements, and 
are computationally expensive to learn, which hinders their application for large scale motion learning and generation. 
Taylor et al. \cite{TaylorNIPS2006} propose a parametric hierarchical modeling of human movements using conditional restricted Boltzmann machines (CRBM) for exact inference and generating stylistic human movements. Higher level layers representing high-level latent factors  can be sequentially added to the model as needed, to capture high-level structure in the movements, resulting in a deep belief network. 
In an extension to \cite{TaylorNIPS2006}, the factored CRBM was introduced that incorporate a set of context/style variables (discrete or continuous) to 
generate different motion styles, synthesize transitions between stylized motions as well as  interpolating and extrapolating the training data~\cite{Taylor:2009}.  In another study, a hierarchical factored CRBM along with a multi-path style interpolation algorithm are used to generate stylistic motions~\cite{Chiu:2011}. in another study to generate stylistic gestures accompanying  speech. However, learning CRBMs is challenging due the large number of parameters and metaparameters that need to be set \cite{HintonRBMGuide}.  


In this paper,  a new example-based affective movement generation approach is presented that aims to modulate a desired motion path to convey a target emotion.  The proposed approach exploits movements from the target emotion class that are kinematically 
similar  to the desired motion path  (nearest neighbours (NN)) 
 to generate a modulated version of the desired motion path that conveys the target emotion. The Laban Effort and Shape abstraction is used to identify the nearest neighbours. 
 The nearest neighbours along with the desired motion path are then encoded in an HMM and the most likely state sequence of the resulting HMM for the desired motion path is used to generate a modulated version of the desired motion path.

\section{Methodology}
\label{sec:method}
The proposed approach  makes use of two movement abstractions: 1) LMA Effort and Shape components, and 2) hidden Markov modeling. Using these abstractions, a new movement is generated that is affectively constrained by a target emotion and kinematically constrained by a desired motion path. 

In this section, the LMA and HMM movement abstractions along with the proposed affective movement generation approach are described.

\subsection{LMA Effort and Shape components}
\label{sec:LabQu}
 The work on affect and its bodily manifestation in psychology and the dance community provides valuable insights that can advance the research in affective computing. In particular, movement notation systems provide a rich tool for a more compact and informative representation of movements, compared to high-dimensional   joint time-series 
trajectories. 

The Laban movement analysis (LMA) is a notation system that provides a set of semantic components based on which both kinematic and expressive characteristics of the movements can be described.
Among LMA components, Effort and Shape are the most relevant for the study of affective movements. Effort describes the inner attitude toward the use of energy and Shape characterizes the bodily form, and its changes in space.  Bartenieff presents Effort and Shape as a complete system
for the objective study of movements,
from behavioural and expressive perspectives \cite{bartenieff1965effort}. Effort has four bipolar semantic components: Weight, Time, Space, Flow (see Table \ref{tab:Effort}), and Shape has three components:
Shape Flow, Directional, and Shaping/Carving\footnote{The Shape components (Shape Flow, Directional, and Shaping/Carving) are referred to as the components of the ''Modes of Shape change" by Peggy Hackney in~\cite{hackney2004making}} (see Table \ref{tab:Shape}).

 In order to enable the application of the LMA Effort and Shape components for computational analysis of affective movements, first, these components need to be quantified. In previous work, we  presented an Effort and Shape quantification method and verified its reliability in comparison with annotations from a certified motion analyst \cite{SamadaniACII2013}. In the following,  the Effort (Weight, Time, and Flow) and Shape Directional quantifications originally presented in~\cite{SamadaniACII2013} and \cite{hachimura2005analysis} are adapted for full-body movements and extended to include quantifications for Shape Flow and Shaping. 
 The Effort Space is not considered  in the present study as the Space describes the focus of attention of the mover (single-focused vs multi-focused)~\cite{dell1977primer} and other visual cues (eye and head movements) might be needed to characterize Space. For instance, an expansive hand and arm movement can be
 used to greet several arriving parties (multi-focused, Indirect) or a single arriving
 person (single-focused, Direct), which would be difficult to annotate without
 additional contextual information. The Space quantification approaches tested in~\cite{SamadaniACII2013} were not in agreement with annotations by a certified motion analyst.  The Space Effort was also excluded in a study using LMA components for motion indexing and retrieval~\cite{motion-retrieval-i3d}.
 
 \begin{table}
  \centering
  \caption{LMA Effort components adapted from \cite{bartenieff1965effort}}
    \begin{tabular}{p{1in}p{.5in}p{1.3in}}
    \hline
    \hline
     & \textbf{Extremes} & \textbf{Example}\\
    \hline
    \textbf{Space:} Attention to   surroundings & \begin{tabular}[t]{@{}p{.5in}p{1.3in}@{}} Direct & Pointing to a particular spot \\ \hline
    Indirect &  Waving away bugs \end{tabular}  & \\\rule{7.6cm}{0.4pt}
    \textbf{Weight:} Sense of the impact of one's movement & \vspace{2pt} \begin{tabular}[t]{@{}p{.5in}p{1.3in}@{}} Light & Dabbing paint on a canvas \\ \hline
       Strong & Punching\end{tabular}  & \\\rule{7.6cm}{0.4pt}
    \textbf{Time:}  Sense of  urgency & 
    \vspace{2pt} \begin{tabular}[t]{@{}p{.5in}p{1.3in}@{}} Sustained & Stroking a pet \\ \hline
     Sudden & Swatting a fly \end{tabular}  & \\\rule{7.6cm}{0.4pt}
    \textbf{Flow:} Attitude toward bodily tension and control  &  \vspace{2pt} \begin{tabular}[t]{@{}p{.5in}p{1.3in}@{}}  Free  & Waving wildly \\ \hline
     Bound & Carefully carrying a cup of hot liquid \end{tabular}  & \\
    \hline
    \end{tabular}%
  \label{tab:Effort}%
\end{table}%

\begin{table*}
  \centering
  \caption{LMA Shape components \cite{hackney2004making}.}
    \begin{tabular}{p{2.5in}p{1in}p{2.6in}}
    \hline
    \hline
    \multicolumn{1}{c}{\textbf{}} & \textbf{Elements} &\textbf{Example} \\
    \hline
    \textbf{Shape Flow:}  is self-referential and defines readjustments of the whole body for internal physical comfort.  & \begin{tabular}[t]{@{}p{1in}p{2.6in}@{}}Growing & Self-to-self communication, stretching to yawn \\ \hline Shrinking & Exhaling with a sigh\end{tabular}  & \\\rule{16.3cm}{0.4pt}
\textbf{Directional:} is goal-oriented and defines the pathway to connect or bridge to a person, object, or location in space.
          & \vspace{5pt}\begin{tabular}[t]{@{}p{1in}p{2.6in}@{}}Arc-like & Swinging the arm forward to shake hands  \\ \hline Spoke-like & Pressing a button \end{tabular} &\\ 
\rule{16.3cm}{0.4pt}          
\textbf{Shaping/Carving:} is process-oriented and is the three dimensional ``sculpting'' of body oriented to creating or experiencing volume in interaction with the environment.  & \vspace{5pt}Molding, \begin{tabular}[t]{@{}cl@{}}\begin{tabular}[t]{@{}p{1in}@{}}Contouring, or \\Accommodating \end{tabular}& Cradling a baby\end{tabular}\\
    \hline
    \end{tabular}%
  \label{tab:Shape}%
\end{table*}%

\subsubsection{Weight Effort}
The  maximum of the sum of the kinetic energy of the torso and distal body limbs (end-effectors: head, hands, feet) is used to estimate the Weight Effort of the whole body. The higher the peak kinetic energy, the Stronger the Weight. 
The sum of the kinetic energy at time $t_i$ is:
\begin{equation}
E(t_i) = E^{\text{Torso}}(t_i) + E^{\text{End-effectors}}(t_i).
\end{equation}
For instance, the kinetic energy of the right hand ($\text{R}_{\text{hand}}$) at time $t_i$ is computed as:
\begin{equation}
E^{\text{R}_{\text{hand}}}(t_i) =\alpha_{\text{R}_{\text{hand}}}{v^{\text{R}_{\text{hand}}}(t_i)}^2,
\end{equation}
where $\alpha_{R_{\text{hand}}}$ is the mass coefficient for the right hand and ${v^{R_{\text{hand}}}(t_i)}^2$ is the square of the speed of the right hand at time $t_i$. The  Weight Effort for a sampled movement of length $T$ (the movement is defined by $T$ sequential observations, where each observation represents Cartesian positions of body joints at a particular $t_i, i \in [1,T]$) is then determined as:
\begin{equation}
Weight = \max_{i\in[1,T]}(E(t_i)).
\end{equation}
In \cite{nakata2002analysis}, the mass coefficients for different body parts of a dancing pet robot were set based
on ``their visual mass and conspicuousness" in the range of 1 to 4 (e.g., $\alpha_{Trunk}$ = 4 and
$\alpha_{Arm}$ = 2). We have set  the mass coefficients to the values proposed in \cite{nakata2002analysis} and observed that the resulting Effort Weight values for different body parts are highly correlated ($>$ 70\%, \textit{p}$<$0.05) with those obtained using the mass coefficients of 1. This observation indicates that the Effort Weight of a body part is mainly influenced by its speed. Therefore, in the present work, mass coefficients are set to 1 for all the body parts as in~\cite{hachimura2005analysis, SamadaniACII2013}. 

\subsubsection{Time Effort}
The weighted sum of the accelerations of the torso and end-effectors is used to estimate the Time Effort for full-body movements. The weights are the mass coefficients similar to the quantified Weight Effort. 
The acceleration\footnote{In the quantifications, the derivatives of motion trajectories used to compute velocity, acceleration, and jerk, are low-pass filtered to remove the high-frequency noise.} for the $k^{th}$ body part at  time $t_i$ is:
\begin{equation}
\label{Eq.Time}
a^k(t_i)=\frac{v^k(t_i)-v^k(t_{i-1})}{\Delta t}.
\end{equation}

Eq.~\ref{Eq.Time} is the discrete form of the differential definition of acceleration, $a=dv/dt$. In this equation, acceleration is defined as the the change in velocity over a unit time ($\Delta t = t_i - t_{i-1} = 1$, where $t~\in~[1, T]$).
Sudden movements are characterized by larger values in the weighted acceleration time-series as compared to Sustained movements.
The maximum of the weighted acceleration time-series is used to describe the  Time Effort of the whole body in the full-body movements.

\subsubsection{Flow Effort}
The Flow Effort for a full-body movement is computed as the aggregated jerk over time for the torso and end-effectors. Jerk is the third order derivative of the position. 
\begin{equation}
Flow = \sum_{\substack{k: \ \text{End-effectors}}} \sum_{\substack{i=2}}^T \frac{|a^k(t_i) - a^k(t_{i-1})|}{t_i - t_{i-1}}, 
\label{FlowEq}
\end{equation}
where $a^k(t_i)$ is the Cartesian acceleration of the $k^{th}$ body part at time $t_i$. 
 
\subsubsection{Shape Shaping}
  Lamb introduces ``Shaping'' as a component used to primarily describe concavity and convexity of the torso in the vertical, horizontal, and sagittal planes \cite{lamb1965posture}. As a result, Shape Shaping defines the processes of Rising/Sinking (vertical plane), Widening/Narrowing (Horizontal plane), and Advancing/Retreating (Sagittal plane)~\cite{lamb1965posture}.
Shaping in the vertical plane is due to torso's upward-downward  displacement~\cite{dell1977primer}; hence, it is quantified as the maximum vertical displacement of the torso.
In the sagittal plane, Shaping is due to the torso's forward-backward displacement~\cite{dell1977primer} and it is quantified as the maximum sagittal  displacement of the torso.

In the horizontal plane (widening/narrowing),  Shaping is  mainly sideward across or out away from the body~\cite{dell1977primer} and  captures the extent of molding of the whole body in the horizontal plane.   In this work, the area of the convex hull encompassing
the body projected onto the horizontal plane is used to capture Shaping in the horizontal plane. This measure captures how far the body traverses  sideward across and out away from the body.

\subsubsection{Shape Flow} 
Cecily Dell suggests the use of the ``reach space" for observing Shape Flow~\cite{dell1977primer}. Three areas of reach are: 1)~near (knitting), 2)~intermediate (gesturing), and 3)~far (the space reached by the whole arm extended out of the body without locomotion). Therefore, the limits of far reach are the limits of what Laban called the personal \textit{kinesphere}, the space around the body which can be reached without taking a step~\cite{dell1977primer}. Given the Shape Flow's relation to the reach space and the limits of personal kinesphere, in this work, the Shape Flow is measured as the maximum of the volume of the convex hull (bounding box) containing the body. 

\subsubsection{Shape Directional}
Since the Shape Directional describes the transverse behaviour of the hand movements (`Spoke-like' or  `Arc-like')~\cite{dell1977primer}, it is captured as  the average curvature of the hand movements in a 2 dimensional (2D) plane within which the largest displacement occurs. There is a separate Shape Directional component for each hand. 
To capture the 2D plane with maximum displacement, we apply multivariate principal component analysis (PCA) on the 3D Cartesian trajectories of the hands
and extract the top two dimensions. The extracted dimensions span the 2D plane where the maximum displacement occurs. 
Next, the average 2D curvature within the PCA-extracted 2D plane ($xy$ plane) for a sampled movement of length $T$ is computed as follows
\begin{equation}
\kappa = \frac{1}{T}\sum_{\substack{i=1}}^T \frac{\sqrt{(\ddot{y}(t_i)\dot{x}(t_i) - \ddot{x}(t_i) \dot{y}(t_i))^2}}{(\dot{x}^2(t_i) + \dot{y}^2(t_i))^{3/2}},
\end{equation}
where $\dot{x}(t_i)$ and $\ddot{x}(t_i)$ indicate the first and second derivatives of the $x$ trajectory at  time $t_i$, respectively. \vspace{1pt}

\subsection{Regularized multinomial logistic regression}
\label{sec:RMRL}
In the proposed approach, a regularized multinomial logistic regression (RMLR) is used to identify affectively non-discriminative LMA components. The multinomial logistic regression uses the inverse logit function to model the posterior of occurrence of different classes in a multi-class problem ($K$ classes) given a set of explanatory variables.
 Given a dataset $\mathcal{D}_p=\{(\mathbf{x}_j,y_j)|\mathbf{x}_j\in\mathbb{R}^{p}, y_j\in\mathcal{Y}, j=1,\dots,N\}$, where $\mathbf{x}_j$ is the $j^{th}$ datapoint of dimensionality $p$ and $\mathcal{Y}$ is the label set for the $K$ classes, the posterior of the $k^{th}$ class is approximated using the symmetric multi-class inverse logit function.

\begin{align}
&Pr(y_i = k| \mathbf{x}_i) = \frac{exp(\mathbf{\beta}_k^\top\mathbf{x}_i)}{1+\sum_{l=1}^{K}exp(\mathbf{\beta}_l^\top\mathbf{x}_i)},\ k=1, \dots, K.
\end{align}
The posterior model parameters for all the classes are denoted by $\theta$
\begin{equation}
\theta = \{\mathbf{\beta}_1^\top, \mathbf{\beta}_2^\top \dots,\mathbf{\beta}_{K}^\top \}, \ \text{where} \ \theta \in \mathbb{R}^{K\times(p+1)}. 
\end{equation}
$\theta$ includes an intercept  along with posterior coefficients for all the classes (posterior coefficient of $k^{th}$ class: $\mathbf{\beta}_k \in \mathbb{R}^p$).  The optimal parameters, $\theta$, are found by solving the maximum log-likelihood problem 

\begin{equation}
 \underset{\mathbf{\theta}}{\argmax} \sum_{i=1}^N \text{log}(P_{y_i}(x_i;\theta)),
\end{equation}
where $P_k(\mathbf{x}_i;\theta) = Pr(y = k| \mathbf{x}_i;\theta)$.

To identify dimensions of $\mathbf{x}_i$ most salient to discriminating between the $K$ classes, the regularized multinomial logistic regression is proposed (RMLR) \cite{Friedman2010}. In this work, the multinomial logistic regression  with elastic net regularization is used, which maximizes the following cost function to find the RMLR model parameters
\begin{equation}
\begin{split}
 \underset{\mathbf{\theta}}{\argmax} \sum_{i=1}^N \text{log} &(P_{y_i}(x_i;\theta))- \dots\\
 & \lambda \sum_{k=1}^K\sum_{j=1}^p(\alpha \|\beta_{kj} \|_1 +(1-\alpha)\|\beta_{kj}\|_2^2).
 \end{split}
 \label{eq:RMLR}
\end{equation} 

$\|.\|_n$ denotes the $l_n$ norm. The tuning variable $\lambda$ controls the strength of regularization; the larger the $\lambda$, the larger the number of class-specific model parameters $\mathbf{\beta}_k$ that will be driven to zero (i.e., fewer LMA components are selected). The variable $\alpha$ is the mixing variable, which controls the contribution of $l_1$ and $l_2$ norms of the model parameters in the regularization ($\alpha = 1$ results in Least Absolute Shrinkage and Selection Operator (LASSO) and $\alpha = 0$ results in ridge regression). The $l_1$ norm imposes sparsity over the model parameters, and the $l_2$ norm encourages the correlated features to be averaged. 
A detailed derivation and solution for multinomial logistic regression can be found in \cite{hastie2009elements}.

\subsection{HMM-based movement modeling}
\label{HMM model}
Hidden Markov modeling is a generative technique that models a sequential observation as a stochastic process whose dynamics are described by a discrete hidden state variable. The hidden state varies between $N_h$ hidden state values based on a state transition matrix $A$ of size $N_h \times N_h$. The observation variables (outputs) are described by a vector of size $n_D$. The distribution of the observations for each hidden state can be modeled as a mixture of $M$ multivariate Gaussians  and is denoted as $\mathfrak{B}$. 
Furthermore, there is an initial state  probability $\pi_i|_{i=1}^{N_h}$ for each hidden state value. Therefore, an HMM  $\lambda$ consists of $\lambda(A,\mathfrak{B},\pi)$.

In the proposed approach, fully-connected hidden Markov models (HMM)s are used. The fully-connected models enable identifying an optimal set of states that best represent a desired motion path without the need to pass through all the states and being restricted with the left-to-right order of the states as is the case in the left-to-right models. 

Efficient algorithms exist for estimating the model parameters $A,\mathfrak{B}, \text{and } \pi$ (e.g., the Baum-Welch algorithm, an expectation-maximization algorithm), evaluating the likelihood that a new observation sequence was generated from the model (e.g., the forward algorithm), and estimating the most likely state sequence (the Viterbi algorithm). A detailed review of HMMs can be found in \cite{RabinerHMM}.
In this work, the Baum-Welch algorithm is used to train HMMs and 
the Viterbi algorithm is used to generate the most likely state sequence for a desired motion path given an HMM of its LMA Effort and Shape nearest neighbours. 

\subsection{The proposed generation approach}
\label{sec:prop}
\begin{figure*}
    \centering
        \includegraphics[trim = 140mm 645 140mm 10mm, scale=0.9]{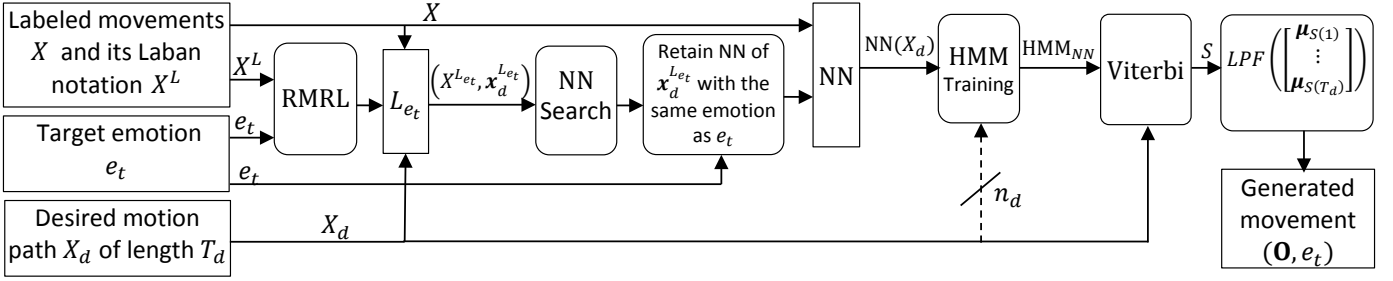}
  \caption{Schematic of the proposed approach. The rectangles represent input data or an outcome of a process and the processes are shown by rounded rectangles. $X^{L_{e_t}}$ and $\mathbf{x}_d^{L_{e_t}}$: the representation of $X$ and $X_d$ in terms of affectively non-discriminative LMA components $L_{e_t} \mid$ NN: indices of nearest neighbours of $X_d \mid$ NN$(X_d)$: nearest neighbours of $X_d$ from $X \mid$ HMM$_{NN}$: HMM of NN$(X_d) \mid S$: the most likely HMM$_{NN}$'s state sequence for $\mathbf{X}_d \mid \mathbf{\mu_i}$ is the mean of the Gaussian output of the $i^{th}$ state of HMM$_{NN} \mid LPF$: low-pass filter $\mid \mathbf{O}$: the modulated $\mathbf{X}_d$ conveying $e_t$.}
\label{fig:diag}
 \end{figure*}

Figure \ref{fig:diag} shows a schematic of the proposed generation approach. 
Consider a sample dataset of $N$ time-series movements $X$, each labelled with one of $M$ affective labels from a set $\mathcal{E}$, $\mathcal{D}=\{(X_j,e_j)|X_j\in\mathbb{R}^{n_D\times T_j}, e_j\in\mathcal{E}, j=1,\dots,N\}$,
where $X_j$ and $e_j$ denote the $j^{th}$ sampled movement of length $T_j$ and its affective label, respectively. 
For a desired motion path $X_d\in\mathbb{R}^{n_D\times T_d}$ and a target emotion  $e_t\in\mathcal{E}$, the proposed generation approach uses movements from the target emotion class $e_t$  that are kinematically 
similar to the desired motion path (viz., kinematic neighbours). 
The identification of  kinematic neighbours of the desired motion path  is necessary to preserve  kinematic characteristics of the motion as much as possible in the process of modulating it to convey the target emotion. 
  $X_d$ may be taken from the dataset $\mathcal{D}$ (e.g., if we wish to take a  movement from the existing set and alter the emotion), or may be specified independently of $\mathcal{D}$. 

A naive approach for identifying kinematic neighbours would be to perform a nearest neighbour search in the high-dimensional joint trajectory space.  However, 
there are two shortcomings associated with the naive approach: 1)~since the movements are time-series observations of different lengths, a direct nearest neighbour search in the space of the Cartesian joint trajectory is not possible, and 2)~in high-dimensional spaces, datapoint pairs are equidistant for a wide range of data distributions and distance metrics~\cite{Beyer1999}.  The second shortcoming is especially problematic as it adversely affects the meaningfulness of the nearest neighbour search in  high-dimensional spaces. To address these shortcomings, in this study, the  search for kinematic neighbours is performed in a low-dimensional space spanned by 
the LMA Effort and Shape components. 
However, LMA Effort and Shape components encode emotions and their intensities in addition to kinematic characteristics of movements~\cite{bartenieff1965effort}. 
As a result, kinematically-similar movements from different emotion classes might occupy distinct regions of the LMA space due to their affective differences 
and finding kinematic neighbours of a movement in such a space might be impossible. 
To overcome this limitation, for a target emotion class, we restrict our nearest neighbour search to kinematically-relevant, affectively non-discriminative  LMA components. 

To identify affectively non-discriminative LMA components, 
 the multinomial logistic regression  with elastic net regularization is used \cite{Friedman2010}, as summarized in Section \ref{sec:RMRL}. The regularized multinomial logistic regression (RMLR) uses  the movements' Effort and Shape components and affective labels as the explanatory variables and class labels, respectively, and weights the Effort and Shape components according to their saliency for each affective class (RMLR assigns zero weights to affectively non-discriminative LMA components for each emotion class).

Given the identified affectively non-discriminative LMA components, the nearest neighbours (NN) of the desired motion path belonging to the target emotion class are next identified. The number of nearest neighbours is set to the number of movements from the target emotion class in the $\epsilon$-neighbourhood of the desired motion path. The $\epsilon$-neighbourhood  is defined as a circle centered at the desired motion path with a radius equal to 10\% of the distance from the desired motion path to the furthest movement from the target emotion class in a space spanned by the affectively non-discriminative LMA components. 

In cases where the desired motion path (the motion path to be modulated in order to convey a target emotion) is affective, its encoded emotion (original emotion)  can be  recognized using a verified automatic affective movement recognition model (e.g., \cite{SamadaniTHMS2014}). Then, for a target emotion, LMA components discriminating between the pair of original and target emotions are identified using RMLR and excluded from the nearest neighbour search as these components contribute to maximum separability between the original and target emotion classes.

After identifying the NN of the desired motion path, they are encoded in an HMM (HMM$_{NN}$). To further preserve the kinematic specificities of a desired motion path in the generated movement, the HMM$_{NN}$ can be augmented with  $n_d$ copies of the desired motion path (shown as dashed line in Figure \ref{fig:diag}). Encoding more copies of the desired motion path  in the HMM$_{NN}$  imposes extra constraints on the generated movement 
favouring stronger adoption of the kinematics of the desired motion path, at the expense of the affective modulation.  Thus, $n_d$ becomes a process parameter which controls the trade-off between strict adherence to either the affective constraint or the kinematic constraint. 
After training the HMM$_{NN}$, the Viterbi algorithm is used to estimate the most likely state sequence for the desired motion path given HMM$_{NN}$. 
A new motion path is generated by concatenating the means of the Gaussian outputs of the HMM$_{NN}$ 
states, following the estimated Viterbi state sequence. 
The resulting  sequence is low-pass filtered to  remove discontinuities that might occur when transitioning to new states \cite{Kulić01072008}. 

The output of the filtering stage is a novel movement kinematically close to the desired motion path and imbued with the target emotion ($\mathbf{O}$ in Figure \ref{fig:diag}).
Kinematic characteristics of the desired motion path are preserved in two stages of the proposed approach: 1) by using an HMM  abstraction of the nearest neighbours  in a space spanned by  affectively non-discriminative LMA components (HMM$_{NN}$), 
 and 2) by using the most likely state sequence of  HMM$_{NN}$ for the desired motion path. Furthermore, in the case where $n_d > 0$, the resulting HMM$_{NN}$ state sequence is further tailored toward kinematic specificities of the desired motion path; hence, the generated movement more closely mimics the kinematics of the desired motion path. The target emotion is overlaid as the movement is generated using an HMM  of the nearest neighbours conveying the target emotion. 
Figure \ref{fig:1} shows the pseudo-code for the proposed approach. 

\begin{figure}
\centering
\begin{boxedcode}
\textbf{Giv}\=\textbf{en:} A  dataset $\mathcal{D}$ with $N$ movements $X$ each \\labeled with one of $M$ affective classes from a set $\mathcal{E}$. \\[5pt]

\underline{\textbf{LMA Effort and Shape representation}}\\
\> $\bullet$ \textbf{Define}  $\mathbf{L}$, a set of all LMA Effort and Shape\\ components (a total of $n_L$ components),\\

\>$\bullet$ \textbf{Compute} $\mathbf{x}_j^L \in \mathbb{R}^{n_L}$: LMA Effort and Shape \\ \> representation, $\mathbf{\forall} \ \text{movements} \ X_j \in \mathcal{D}$,\\

\> \quad \quad \quad {$\mathcal{D}^L = \{\mathbf{x}_1^L, \dots,\mathbf{x}_N^L\}$},\\[5pt]

\underline{\textbf{Regularized multinomial logistic regression}:}\\
\> $\bullet$ \textbf{Input:} $\{(\mathbf{x}_j^L,e_j)| \mathbf{x}_j^L \in \mathcal{D}^L, e_j \in \mathcal{E}\}$,\\
\> $\bullet$ \textbf{Out}\=\textbf{put:} the  set of affectively discriminative \\
\>LMA components  for each class $k$ ($\mathbf{L}'_k$).\\ 
 \>\> $\bullet \ \mathbf{L}_k = \mathbf{L} \setminus   \mathbf{L}'_k$, affectively non-discriminative \\
 \>\>LMA components for $k^{th}$  class.\\ [5pt]
 
\underline{\textbf{Affective movement generation}}\\
 \> $\bullet$ \textbf{Given} 
  $\begin{cases}
    X_d,  & \text{a desired motion path of length} \ T_d, \\
    e_{t}, & \text{a target emotion},
  \end{cases}$\\[3pt]
  
\>$\bullet$ \textbf{Let} $\mathbf{L}_{e_t}$: a set of affectively non-discriminative\\ 
\> LMA components for $e_{t}$, \\
 
\> $\bullet$ \textbf{Find}\= \ nearest neighbours of $X_d$ in the space\\
\> spanned by $\mathbf{L}_{e_t}$ $\rightarrow \text{NN}^{L_{e_t}}(X_d)$, \\
\> \> -- Retain those that belong to $e_{t}$:\\
\quad \=NN$(X_d)= \{(X,e) \in \mathcal{D}| X \in \text{NN}^{L_{e_t}}(X_d), e = e_{t}\}$,\\ [5pt]
\> $\bullet$ \textbf{H}\=\textbf{MM Training}\\
\>\> -- Train HMM$_{NN}$ of $N_h$ states to model NN$(X_d)$\\
\>\>    along with $n_d$ copies of $X_d$; $n_d \geq 0$. \\[5pt]

\> $\bullet$ \textbf{Viterbi algorithm using HMM$_{NN}$ and $X_d$} \\
	\> \> S$\leftarrow$ most likely state sequence for $X_d$\\
	\> \> $S = [S_1, \dots, S_i,\dots, S_{T_d}], S_i \in \{1,\dots, N_h\}$,\\

\> $\bullet$ \textbf{Motion generation}\\
	\> \> $\textbf{O} = [ \mathbf{\mu}_{S_1}, \mathbf{\mu}_{S_2}, \dots, \mathbf{\mu}_{S_{T_d}}]^\ast$ \\ [3pt]
	\>  \> \textbf{O} $\leftarrow$ \textbf{Low-pass Filter}(\textbf{O}) \\[5pt]
\underline{\textbf{Return:}} \\
\> \textbf{O:} A modulated version of $X_d$ conveying $e_{t}$.	\\
\rule{8.2cm}{0.02cm}\\
\scriptsize  $\ast \mathbf{\mu_i}$ is the mean of the Gaussian output of the $i^{th}$ state of HMM$_{NN}$.
\end{boxedcode}
\caption{Pseudo-code for the proposed generation approach.}
\label{fig:1}
\end{figure}

\section{Experimental setup}
\label{sec:exp}
A publicly available  full-body 
movement dataset~\cite{kleinsmith2006cross} is used to demonstrate the performance of the proposed approach. A 10-fold division is used to divide the dataset into testing and training movements. 
In each fold, a testing movement is considered the desired motion path and the proposed method is applied to produce a new movement conveying the target emotion. Note that in these experiments, the testing movement already contains affective content.  The target emotion is selected to be different than the original emotion encoded in the testing movement.
 The training movements in each fold establish the labeled dataset required for the proposed approach.  The affective quality of the generated movements is objectively evaluated using a state-of-the-art recognition model \cite{SamadaniTHMS2014}. Furthermore, a subjective evaluation is conducted via a user study in which participants rate the affective qualities of the generated movements. 

\subsection{Full-body affective movement dataset}
\label{sec:dataset}
\begin{figure*}
\centering
\includegraphics[trim = 0 35  0 10,scale=1]{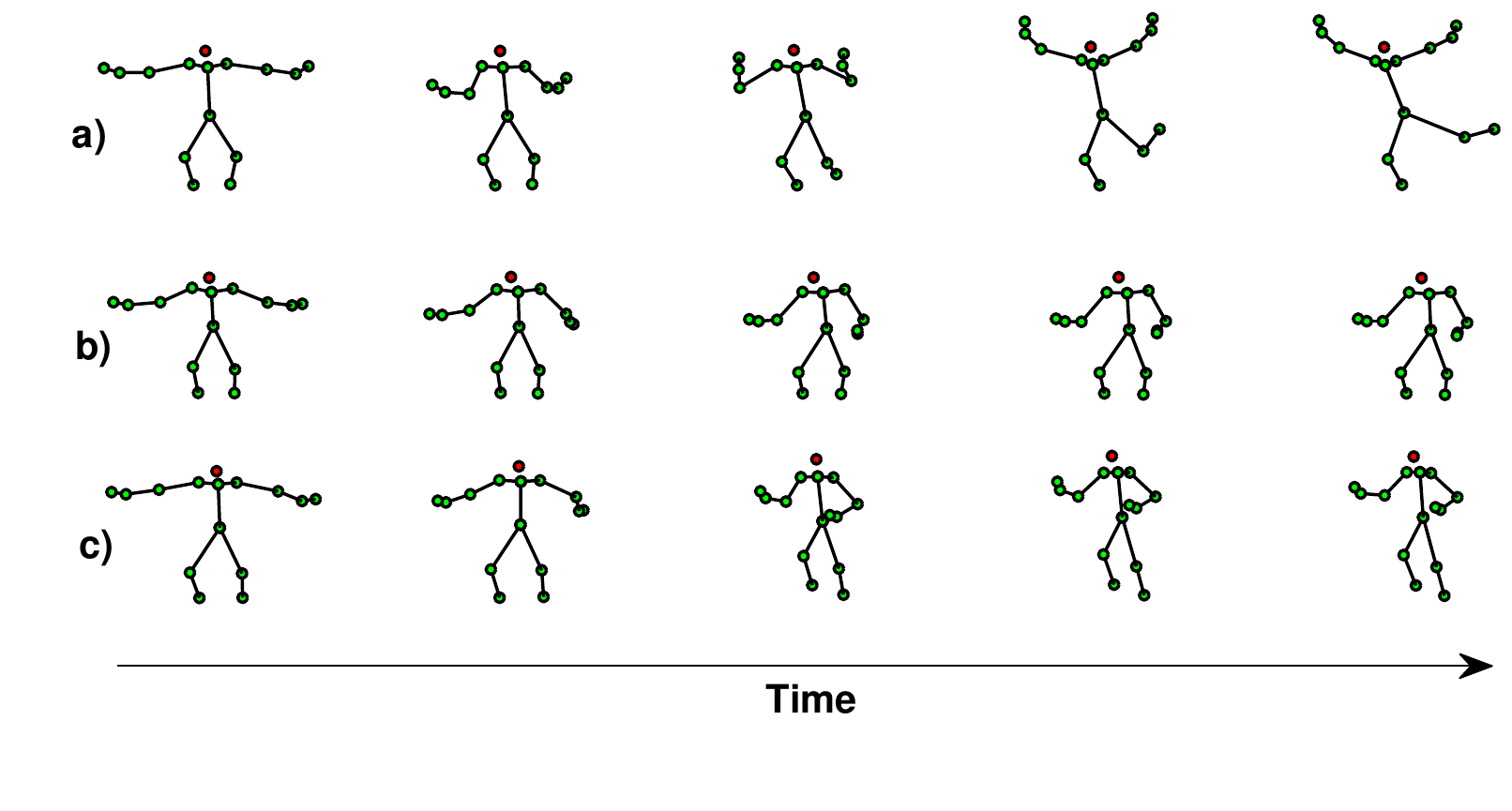}
\caption{Movement exemplars from the full-body dataset: (a) and (b) are two kinematically different happy movements, whereas the happy movement in (b) is similar to the fear movement in (c).}
\label{FBData}
\end{figure*}

The full-body dataset~\cite{kleinsmith2006cross} was collected using a Vicon 612 motion capture system and contains 183 acted full-body affective movements obtained from thirteen demonstrators who freely expressed movements conveying anger, happiness, fear, and sadness with no kinematic constraints.
There are 32 markers attached to bodily landmarks\footnote{The markers are placed on the following bodily landmarks: left front head, right frond head, left back head, right back head, top chest, center chest, left front waist, right front waist, left back waist, right back waist, top of spine, middle of back, left outer metatarsal, right outer metatarsal, left toe, right toe, left shoulder, right shoulder, left outer elbow, right outer elbow, left hand, right hand, left wrist inner near thumb, right wrist inner near thumb, left wrist outer opposite thumb, right wrist outer opposite thumb, left knee, right knee, left ankle, right ankle, left heel, right heel.} and their 3 dimensional Cartesian coordinates are collected using 8 motion capture cameras at a rate of 120Hz\footnote{Some movements in the full-body dataset are recorded at 250Hz, which were down-sampled to 120Hz in our study to keep the sampling rate consistent for all the movements}.
There are 46 sad, 47 happy, 49 fearful, and 41 angry movements in the full-body dataset. Each movement starts from a known pose, the T-pose (arms raised to shoulder level and extended to the sides). The full-body movements range from 1.39 seconds (167 frames) to 9.17 seconds (1100 frames)  in duration.
In  a  user study where participants rated 
the most expressive postures of the movements~\cite{kleinsmith2006cross},  the intended emotions were rated the highest among other emotions. 
 Due to interpersonal (and possibly idiosyncratic) differences as well as the lack of kinematic constraints on the movements, there is a wide range of within-class 
variabilities (Figure \ref{FBData}a and \ref{FBData}b). There are also between-class kinematic similarities (Figure \ref{FBData}b and \ref{FBData}c). The kinematically-unconstrained movements of the UCLIC are a favourable property of this dataset as it allows us to examine the performance of the proposed approach in the presence of interpersonal, between- and within-emotion kinematic differences in the bodily expression of affect. 

The expressivity of the UCLIC dataset has been evaluated in previous a user study in which participants rated intended emotions among other emotions for all the movements in the dataset~\cite{kleinsmith2006cross}. In that user study, an average  recognition rate of 54.7\% was reported  with the least recognized emotion being fear
ones (49.4\% average recognition rate), and the most recognized being  sadness  (63.4\% average recognition rate).  

\subsection{RMLR parameter selection}
In order to identify affectively non-discriminative LMA components for each target emotion,  multinomial logistic regression with elastic net regularization RMLR is used. To set the tuning parameters of RMLR, $\alpha$ and $\lambda$ in Equation \ref{eq:RMLR}, a two-dimensional cross-validation is used.  $\alpha$ values in the range of $0.05:0.05:1$ and 100 values of $\lambda$ are tested and the pair of $\alpha$ and $\lambda$ with the minimum 10-fold cross-validation error (misclassification rate) is selected\footnote{The generalized linear models toolbox from \cite{friedmanglmnet} was used to perform RMLR}.

\subsection{HMM Initialization and model selection}
In order to reduce variability due to  differences in demonstrators' physical builds (e.g., height), each movement in the dataset is normalized by the length of its demonstrator's hand and arm measured from his/her right shoulder to the right wrist.  

Since all the movements in the full-body dataset start from a known T-pose, the hidden state sequence for the HMMs always starts at state 1; hence, the hidden state priors are: $\pi_{i} = 1$ for $i = 1$, and  $\pi_{i} = 0$, otherwise.
The full-body movements are generally very short and progress from the start toward the end with instances of cyclic movements. In this work, all transitions between   states were allowed to capture the cyclic behaviours. Using fully-connected HMMs, the Viterbi algorithm identifies a sequence of HMM states that best represents a desired motion path without the need to pass through all the HMM states, sequentially. 

As described in Section \ref{sec:method}, movements that are kinematically similar to a desired motion path along with $n_d$ copies of the desired motion path are encoded in an HMM (HMM$_{NN}$). In the experimental evaluation of the proposed approach, no copies of the desired motion path are included in the HMM$_{NN}$ abstractions ($n_d = 0$), favouring the affective constraint over the kinematic constraint.  
In the HMM$_{NN}$ abstractions, the distribution of the  movements for each hidden state is modeled as a single Gaussian output.
Generally, for generation purposes, an HMM with a large number of hidden states is favoured to capture fine details of the motion path. We tested different numbers of states ranging between 10-14 states and chose 12 states by a visual inspection for the quality of the generated trajectories. 

To initialize the HMM training process, the identified nearest neighbour movements for a desired motion path are divided into $N_h$ (number of hidden states) equal segments and their means and covariances are used as the initial means and covariances of the output Gaussians associated with the hidden states. Full covariances are used to account for  potential correlations between body joint trajectories.  

\subsection{Affective movement validation}
To evaluate the performance of the proposed approach, the expressivity of the generated movements is assessed objectively using an automatic recognition model, and subjectively via a user study.

\subsubsection{Objective validation}
The automatic recognition approach \cite{SamadaniTHMS2014} is based on a hybrid generative-discriminative modeling of movements and is shown to achieve high interpersonal recognition rates\footnote{In an interpersonal recognition model, testing and training movements come from different individuals}. The recognition model derives a Fisher score representation of the generated movements based on class-specific HMMs and performs nearest neighbour classification in a space spanned by features maximally dependant on the affective categories and encompassing the main modes of variations in the affective movements \cite{SamadaniTHMS2014}. 
 Note that none of the generated movements in this study were used to train the recognition model.

\subsubsection{Subjective validation}
\label{sec:subjective}
Since affective movement perception is  subjective, we have also evaluated the expressivity of the generated movements with a questionnaire-based user study. The user study aims to determine whether the target emotion is perceived from a generated movement, the generated movement is confused as conveying other emotions, or no emotion is perceived from the generated movement. A questionnaire similar to the one in a previous study \cite{SamadaniSocial2013} is designed in which participants are asked to evaluate the expressivity of the generated affective movements in terms of the six basic Ekman emotions (anger, sadness, happiness, fear, disgust, and  surprise) on a 5-point Likert scale ranging between ``0: not conveyed at all'' to ``4: strongly conveyed''. 
 
Using the 10-fold division of the full-body dataset into testing and training movements, a large number of  movements (549 movements) are generated and evaluating the subjective perception of all these movements is not feasible. 
To reduce the number of movements for evaluation, we identify and remove movements which are similar to others, keeping only exemplars for evaluation.
To identify similar movements in a conversion class, the generated movements from the conversion class  are represented in terms of LMA Effort and Shape components and \textit{k}-means clustered. Different numbers of clusters (2 to 5) are tested and the best number of clusters is selected to maximize the goodness of clustering metric \cite{pkekalska2005dissimilarity} defined as 
\begin{equation}
GOC =\frac{\sum_{i=1}^k n_i\sum_{j\neq i} \frac{n_i}{n-n_i}{d}_{ij}}{2\sum_{i=1}^kn_i{d}_{ii}},
\label{GOC}
\end{equation} 
where ${d}_{ij}$ and ${d}_{ii}$ are the average distances between cluster $i$ and $j$ and within cluster $i$, respectively.

Next, the popularity of the identified clusters for each conversion class is inspected and the closest movement to the center of the most populous cluster is selected as the exemplar of that  conversion class for evaluation in the user study. Therefore, there are 12 generated movements used in the user study, each representing a conversion class. For the conversion class of `sadness to happiness', there are two populous clusters with an equal number of members. In this case, the cluster whose representative displays a more kinematically distinct movement than the other generated movements was selected. This is done solely to include a wider range of kinematic movements in the user study. 
For each movement, participants were asked to rate the six basic Ekman emotions, each on 
a 5 point Likert scale. To test whether the target emotion is unambiguously recognized, the participants in the user study were asked to rate each basic emotion separately, and were not forced to chose between emotions; they could indicate that they observed more than one emotion, or that they observed no emotional content. The movements were presented to the participants in a randomized order.

A pilot user study was first conducted with 6 participants to obtain an estimate of effects of the original and target emotions on the participants' perception. The estimated effects vary in size and range from very small effects (e.g., $\eta^2 = 0.001$)  to large ones (e.g., $\eta^2 = 0.745$). Using the G$^*$Power software \cite{faul2007g}, the sample size required to detect a potential significance of the estimated effects of medium size ($\eta^2 = 0.06$; as recommended by Jacob Cohen \cite{cohen2013statistical}) or larger at the statistical power of 90\% is found to be 16 participants. 

17 participants (11 male and 6 female) completed the questionnaire. Participants were healthy adults and recruited from among the students at the University of Waterloo via email. They were provided with the detailed information about the study and the procedure to complete the questionnaire. The study received ethics approval from the Office of Research Ethics, University of Waterloo, and a consent form was signed electronically by each participant prior to the start of the questionnaire.

Participants' responses are then analyzed to assess whether the generated movements communicate the target emotions and if there exists any confusion between the basic emotions.

\section{Results}
\label{sec:res}
As described in Section \ref{sec:dataset}, the dataset used for the experimental evaluation of the proposed approach contains affective full-body movements for 4 emotion classes: sadness, happiness, fear, and anger. Following the proposed approach, each testing movement is converted to convey three emotions other than its own, resulting in 12 conversion classes. The following naming convention is used hereafter to refer to the conversion classes: `original to target' (e.g., `sadness to happiness' indicate the conversion from the original emotion `sadness' to the target emotion `happiness'). 

As described in Section \ref{sec:exp}, a 10-fold division is used to divide the affective full-body movement dataset into testing and training movements. In the experimental evaluations, the testing movements are the desired motion paths, and the training movements form the labeled dataset $\mathcal{D}$ used in the proposed approach as described in Section \ref{sec:prop}.  

Using the quantification described in Section \ref{sec:LabQu}, movements are annotated in terms of Effort components (Weight, Time, and Flow) and Shape components (Shaping, Directional, and Flow). For the Effort components, in addition to the whole-body annotation, these components are also computed for individual body parts: head, right hand, left hand, right foot, and left foot. 
Since, all the movements (testing and training) used in the experimental evaluation are affective (intended to  convey a specific affect), the LMA components that discriminate between emotions encoded in the training movements and those encoded in the testing movement are first identified using RMLR, and excluded from the search for the nearest training movements. 
This will allow finding movements that are kinematically most similar to a desired motion path  from within the target emotion  class (training movements in the $\epsilon$-neighbourhood of the desired motion path that belong to the target emotion class). 
Furthermore, this will result in a computationally efficient nearest neighbour search based on a few LMA components. 

We denote the set of all LMA Effort and Shape components as $\mathbf{L}$, then the set of affectively non-discriminative LMA components for a pair of emotions $(e_1, e_2)$, $\mathbf{L}_t$, is defined as
\begin{equation}
\mathbf{L}_t=\mathbf{L} \setminus (\mathbf{L}_{1} \cap \mathbf{L}_2),
\label{eq:non_dis}
\end{equation}
where $\mathbf{L}_{1}$, and $\mathbf{L}_2$ are the  LMA Effort and Shape components salient to $e_1$ and $e_2$, respectively. 
Table \ref{tab:ComonSig} shows the discriminative LMA components between each pair of emotions in the full-body dataset. 

\begin{table}[htbp]
  \centering
  \caption{Discriminative LMA Effort and Shape components for each pair of emotions identified using regularized multinomial logistic regression.}
    \begin{tabular}{lp{2.1in}}
    \hline
          & \textbf{LMA Effort and Shape components}\\
    \hline
    \textbf{Sadness-Happiness}  &  WeightRFoot, TimeAll, TimeTorso, TimeRHand, TimeLFoot, FlowRFoot, FlowLFoot, ShapeHor, ShapeFlow, ShapeDirRHand \\ [3pt]
    \textbf{Sadness-Fear} & TimeTorso, TimeRHand, TimeLFoot, TimeHead, FlowRFoot, ShapeZ, ShapeHor, ShapeFlow \\ [3pt]
    \textbf{Sadness-Anger} & WeightRFoot, TimeTorso, TimeRHand, TimeRFoot, TimeLFoot, ShapeHor, ShapeDirRHand\\ [3pt]
    \textbf{Happiness-Fear} &  WeightTorso, TimeRHand, TimeLFoot, FlowHead, ShapeZ, ShapeDirRHand\\ [3pt]
\textbf{Happiness-Anger }& TimeLHand, FlowLFoot, FlowHead, ShapeHor, ShapeFlow
 \\ [3pt]
\textbf{Fear-Anger }& WeightHead, TimeTorso, TimeLFoot, TimeHead, ShapeZ, ShapeHor\\
    \hline
    \end{tabular}%
  \label{tab:ComonSig}%
\\  \scriptsize R:right, L:left, Hor: Horizontal, Dir: Directional. 
\end{table}%

Among the emotion classes in the full-body dataset, sadness and happiness have the largest number of LMA components discriminating between them (see Table \ref{tab:ComonSig}), which could be due to the distinct natures of these two emotions. For instance, in the circumplex space \cite{Russell1980}, sadness and happiness are located at opposing extrema of the arousal and valence dimensions.   Shape Horizontal and Time for the left foot are the most frequent discriminative LMA components presented in 5 out of 6 cases in Table \ref{tab:ComonSig}. 
There are components that are found discriminative only for one pair of emotions: 1) TimeAll measures the Effort Time for the whole body and is found to be discriminative for the sadness-happiness pair. 2) Weight Effort for torso (WeightTorso) is found to be discriminative for the happiness-fear pair. 
3) Time for the left hand (TimeLHand) is discriminative for the happiness-anger pair, and 4) Weight for the head (WeightHead) discriminates between fear and anger. It should be emphasized that the identified LMA components in Table \ref{tab:ComonSig} collectively contribute to maximum separability between the pairs of emotions and when considered individually may not result in maximum separation.    

Next, the proposed approach is used to generate movements that convey target emotions. 
In the experimental evaluation presented in this section, no copies of the desired motion path are included in the HMM$_{NN}$ abstractions ($n_d = 0$).  Videos of movement exemplars generated by alternative implementations of the proposed approach ($n_d=0$ and $n_d=1$) are included as supplementary materials.

\subsection{Objective validation}
Table \ref{tab:FishRecog} shows the confusion matrix for the automatic recognition of generated full-body movements.

\begin{table}[htbp]
  \centering
  \caption{Confusion matrix (\%) for the automatic recognition of the generated movements}
    \begin{tabular}{rr|rcccc}
    \hline
          & \multicolumn{1}{r}{} &       & \multicolumn{4}{c}{Recognized Emotion} \\
\cline{4-7}          & \multicolumn{1}{r}{} &       & Sadness & Happiness   & Fear & Anger \\
    \hline
    \multicolumn{1}{c}{\multirow{4}[2]{*}{\begin{sideways}Target\end{sideways}}} & \multicolumn{1}{c|}{\multirow{4}[2]{*}{\begin{sideways}Emotions\end{sideways}}} & \multicolumn{1}{l}{Sadness} & \textbf{83}    & 1     & 15    & 1 \\
    \multicolumn{1}{c}{} & \multicolumn{1}{c|}{} & \multicolumn{1}{l}{Happiness} & 1     & \textbf{61}    & 22    & 15 \\
    \multicolumn{1}{c}{} & \multicolumn{1}{c|}{} & \multicolumn{1}{l}{Fear} & 3     & 8     & \textbf{77}    & 13 \\
    \multicolumn{1}{c}{} & \multicolumn{1}{c|}{} & \multicolumn{1}{l}{Anger} & 1     & 15    & 16    & \textbf{67} \\
    \hline
    \end{tabular}%
  \label{tab:FishRecog}%
\end{table}%

The generated affective movements are recognized above chance with a 72\% recognition rate, which is comparable with the 72\% interpersonal recognition rate reported in \cite{SamadaniTHMS2014}. Similar confusions between emotion classes as those in \cite{SamadaniTHMS2014} are observed here. 
Nevertheless, the automatic recognition rate of the generated movements is high and demonstrates the congruence of the generated emotions with the target ones. 

\subsection{Subjective validation} 
\label{sec:sub}
The performance of the proposed approach  is also evaluated subjectively via a questionnaire-based user study as described in Section \ref{sec:subjective}. 
 Using a one-way ANOVA, the effect of the order in which the videos were presented to a participant was not found significant at $p<0.05$, indicating no habituation was observed as a result of the presentation order.  
 
 To inspect the effects of the original and target emotions on the participants' perception of the generated movements, a mixed-model ANOVA with repeated measures was performed for each emotion rating. In the ANOVA tests, the original and target emotions were the independent variables and participant ratings of the basic emotions were the dependant variables. Initially, the participant's gender was also included as a between-subject variable, however, its effect was not found significant at $p<0.05$, thus, the ANOVA tests were repeated without the gender variable. 
 The SPSS statistical software package was used for the analysis.

\begin{figure*}
 \centering
         \includegraphics[trim= 0 0 0 0, scale=0.75]{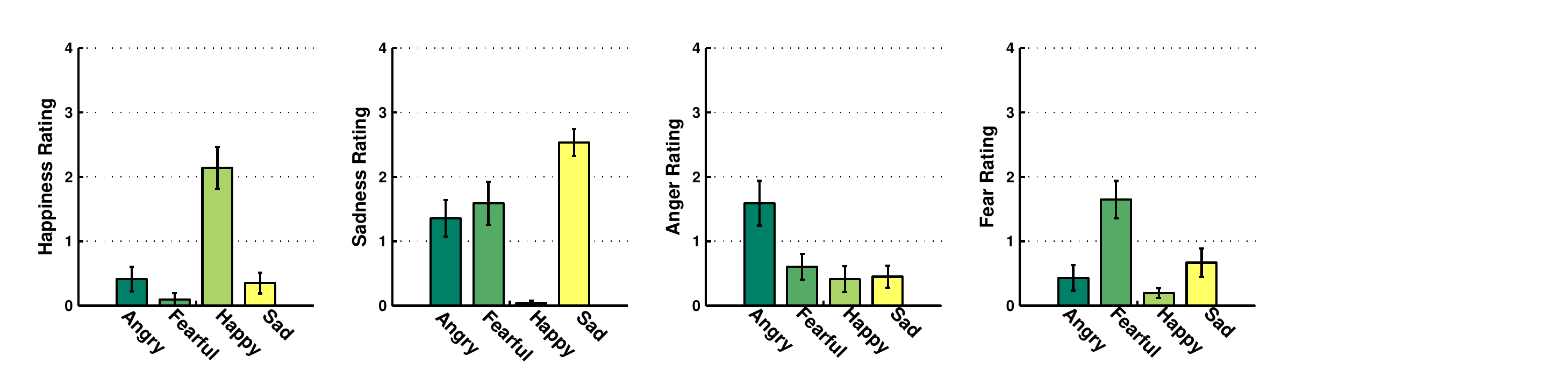}
        \caption{Average participants' ratings (mean$\pm$SE) for the target emotions. Each bar shows an average rating across all the movements with the same target emotion (x-axis is the target emotion). }
\label{fig:rating}
\end{figure*}

It was found that the target emotion has a significant main effect in all  cases ($p<0.005$), and original emotion has a main effect on participants' rating of anger, sadness, fear, surprise ($p<0.05$). There are also significant interaction effects between the original and target emotions in all  cases except for the rating of disgust.
 Furthermore, participants' ratings of an emotion were found to be significantly higher 
 (at a Bonferroni corrected significance level of $p<0.0125$ with an effect size Cohen's $d > 0.67$) when the emotion is the target emotion (e.g., sadness rating for anger to sadness conversion) as compared with cases when the emotion is not the target emotion (e.g., sadness rating in anger to happiness conversion). This observation can also be seen in the average participants' ratings of target anger, sadness, fear, and happiness  emotions shown in Figure \ref{fig:rating}.

The detected main effects of the target emotion indicate that the proposed approach successfully encoded the target emotion in the generated movements. 
The detected significant effects of the original emotion might indicate the presence of residues of the original emotion in the generated movements or may be an indicator of deficiencies  in the perception of emotion from body movements. 

\begin{figure*}
 \centering
         \includegraphics[trim=0 15mm 0 0 0, scale=0.85]{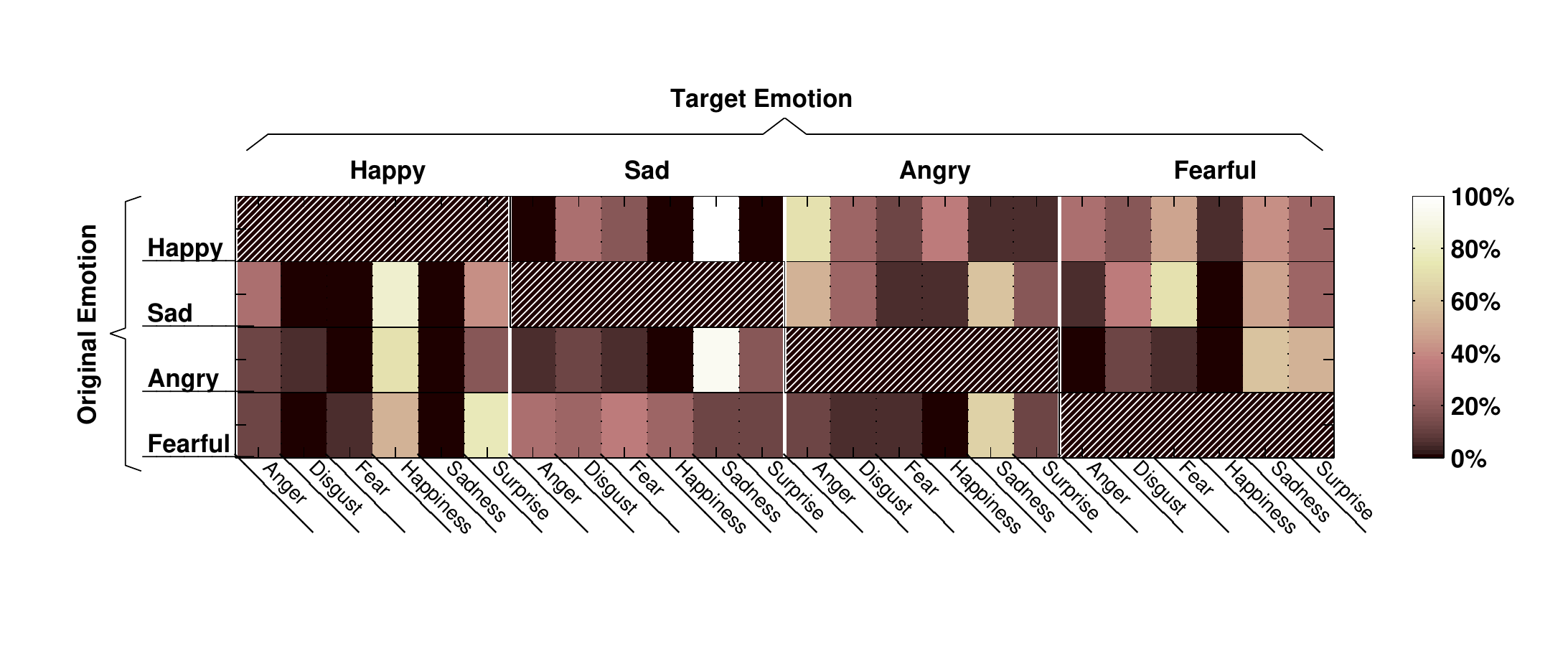}
        \caption{A heatmap showing the average recognition rates of different emotions for the conversion classes.  An emotion is considered recognized if it is rated 2 or above on the Likert scale spanning between 0 to 4. Note that this recognition cut-off is applied for illustrative purposes in this figure only and all the subjective evaluation reported in this section is performed on the full scale of rating. }
\label{fig:Heat}
\end{figure*}

\begin{figure*}
 \centering
         \includegraphics[trim=0 15mm 0 0 0, scale=0.85]{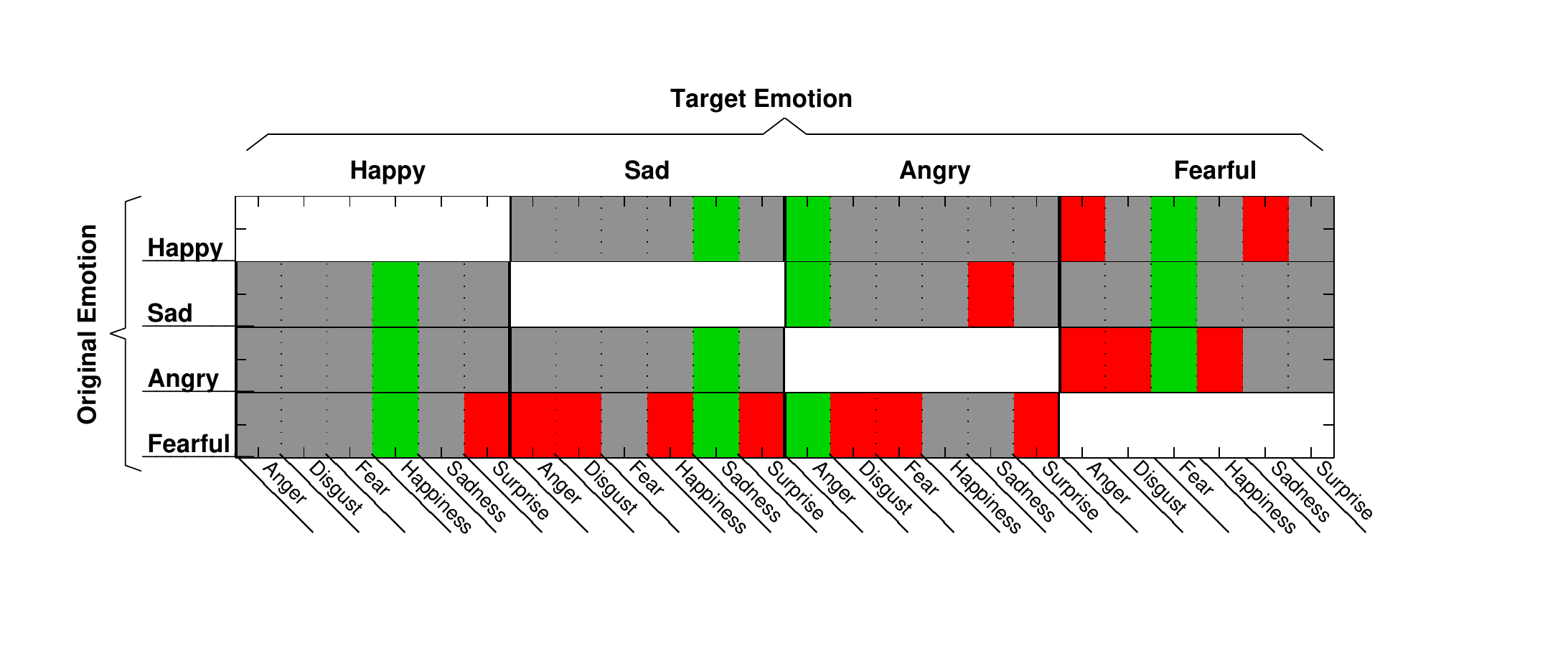}
        \caption{A heatmap showing significance of pair-wise differences between participants' ratings of target emotions and other emotions (paired \textit{t}-tests). The green boxes highlight the target emotion, the grey boxes indicate significant differences to the ratings of the target emotion at $p<0.05$, and a red box indicates that there is no significant difference to ratings of the target emotion at $p<0.05$. }
\label{fig:Sig}
\end{figure*}

A closer look at the participants' perception of individual movements reveals a few cases of confusion between the emotions, some of which are between the original and target emotions.  
Figure \ref{fig:Heat} is a heat map showing the average recognition of different basic emotions for each conversion class (confusion matrix). To explore significant pair-wise differences, for each movement, paired \textit{t}-tests between participants' ratings of its target emotion and those of other emotions were performed. Figure \ref{fig:Sig} highlights the results of the paired \textit{t}-tests. The green boxes are the target emotions, grey boxes are the emotions whose ratings are significantly different than the corresponding target emotion at $p<0.05$, and red boxes are those emotions whose ratings are not significantly different than the target emotion at $p<0.05$.
The Cohen's $d$ effect size was also computed to evaluate the size of pair-wise differences between participants' ratings of target emotions and those of other emotions for each movement.  Cohen's recommendation for effect size evaluation is used here (small: $d<0.2$, medium: $d = 0.5$, and large: $d>0.8$). 

 As can be seen in Figures \ref{fig:Heat} and \ref{fig:Sig},
  the target emotion is clearly perceived in the conversions from anger and sadness to happiness at a rate $>$71\% (significant pair-wise differences are detected between ratings of happiness and other emotions at $p<0.001$ with effect sizes $d>1.22$), whereas in the fear to happiness conversion, there is  no significant difference between ratings of happiness and surprise  
  ($p = 0.13$ and small to medium effect size of $d = 0.41$) and both happiness and surprise are rated high.  
  
 When the target emotion is sadness, it is correctly perceived for the conversions of happiness and anger to sadness at a rate $>$94\%, $p<0.001$, and $d >2.76$ (2$^{nd}$ column of Figures \ref{fig:Heat} and \ref{fig:Sig}). For the fear to sadness conversion, the target emotion is only recognized at a 12\% rate and no other emotion is rated high.

Anger is correctly perceived for the movement converted from happiness at a rate of 71\% with significant pair-wise differences between ratings of the anger and other emotions at $p<0.028$ with effect sizes $d>0.93$ (3$^{rd}$ column of Figures \ref{fig:Heat} and \ref{fig:Sig}).
For the sadness to anger conversion, anger and sadness are rated the highest  with a small non-significant  pairwise difference ($p = 0.47, d = 0.035$) between the ratings of anger (recognition rate of 53\%) and those of sadness (recognition rate of 59\%). This could have been caused by residues of sadness remaining in the converted movement (main effect of the original emotion on anger ratings observed in the ANOVA test). 
For the fear to anger conversion, 
the generated movement is confused as a sad movement with a recognition rate of 65\% (Figure \ref{fig:Heat}). 

When the target emotion is fear, it is correctly recognized when converted from sadness at a rate of 71\% with significant pair-wise differences between ratings of fear and those of other emotions at $p<0.01, d>1.07$. For happiness to fear conversion, despite the highest rating for fear among other emotions (recognition rate of 47\%), it is not significantly different than the ratings of sadness ($p = 0.46, d = 0.04$) and anger ($p= 0.17, d= 0.34$) (Figure \ref{fig:Sig}). The fearful movement converted from anger is confused as sad (recognition rate of 59\%) and surprise (recognition rate of 53\%) (Figure \ref{fig:Heat}).

\section{Discussion}
\label{sec:dis}

The performance of the proposed approach in generating movements with recognizable target emotions was evaluated objectively using an automatic recognition model from \cite{SamadaniTHMS2014}, and subjectively via a user study. 

The encoded emotions were correctly recognized by the automatic recognition model from \cite{SamadaniTHMS2014} for 72\% of the generated movements, which is comparable with the recognition rate achieved by the model for the affective movements from 13 demonstrators~\cite{SamadaniTHMS2014}.
 The observed confusions reported in Table \ref{tab:FishRecog} are mainly between fear, happiness, and anger. Considering the circumplex model of emotion~\cite{Russell1980}, the observed confusions seem to be related to the similarities between the affective expressions along the arousal and valence dimensions. For instance, anger and fear are both high arousal and negative valence expressions, and they were frequently confused in the objective validation experiment.

In the subjective validation, the effect of the target emotion on the participants' perception was found significant. However, there are cases where the target emotion was confused with other emotions. In all of these cases, either the original or target emotion is fear with the exception of the sadness to anger conversion. 
One hypothesis is that the observed confusions could have  resulted from inaccurate training movements. The movements in the full-body dataset are labeled by their demonstrators, and therefore, they might not fully communicate the demonstrator-intended emotions. 
Inclusion of these movements would degrade the performance of the proposed approach as the training movements used to modulate a desired motion path to convey a target emotion might  not be sufficiently (or accurately) affective.  
Therefore, the performance of the proposed approach depends on the availability of a dataset containing a diverse range of training movements whose expressivity is verified by different observers. 
In addition, the observed deficiencies in the perception of fear are congruent with reports from the literature \cite{van2007body, montepare1987,+Coulson2004}, which indicate that other modalities in addition to body movements might be required for the perception of fear.
After excluding the cases with original or target fear emotion, the participants recognized the target emotions in the generated movements at a rate of 71\% in the user study. 

A direct comparison between the recognition rates from the objective and subjective evaluation experiments is not possible as the objective evaluation was performed for all the generated movements, whereas the subjective evaluation was done for twelve movement exemplars only (one for each conversion class).  
Nevertheless, the higher objective recognition rates (especially for generated fearful movements) are likely because the automatic recognition model was  trained using the affective movements in the full-body dataset to discriminate between four emotions (anger, happiness, fear, and sadness), while in the user study, participants rated all the six basic Ekman emotions for each movement, which might have rendered the rating of target emotions more difficult. 

\begin{figure}
 \centering
         \includegraphics[trim=0 0mm 0 0 0, scale=0.75]{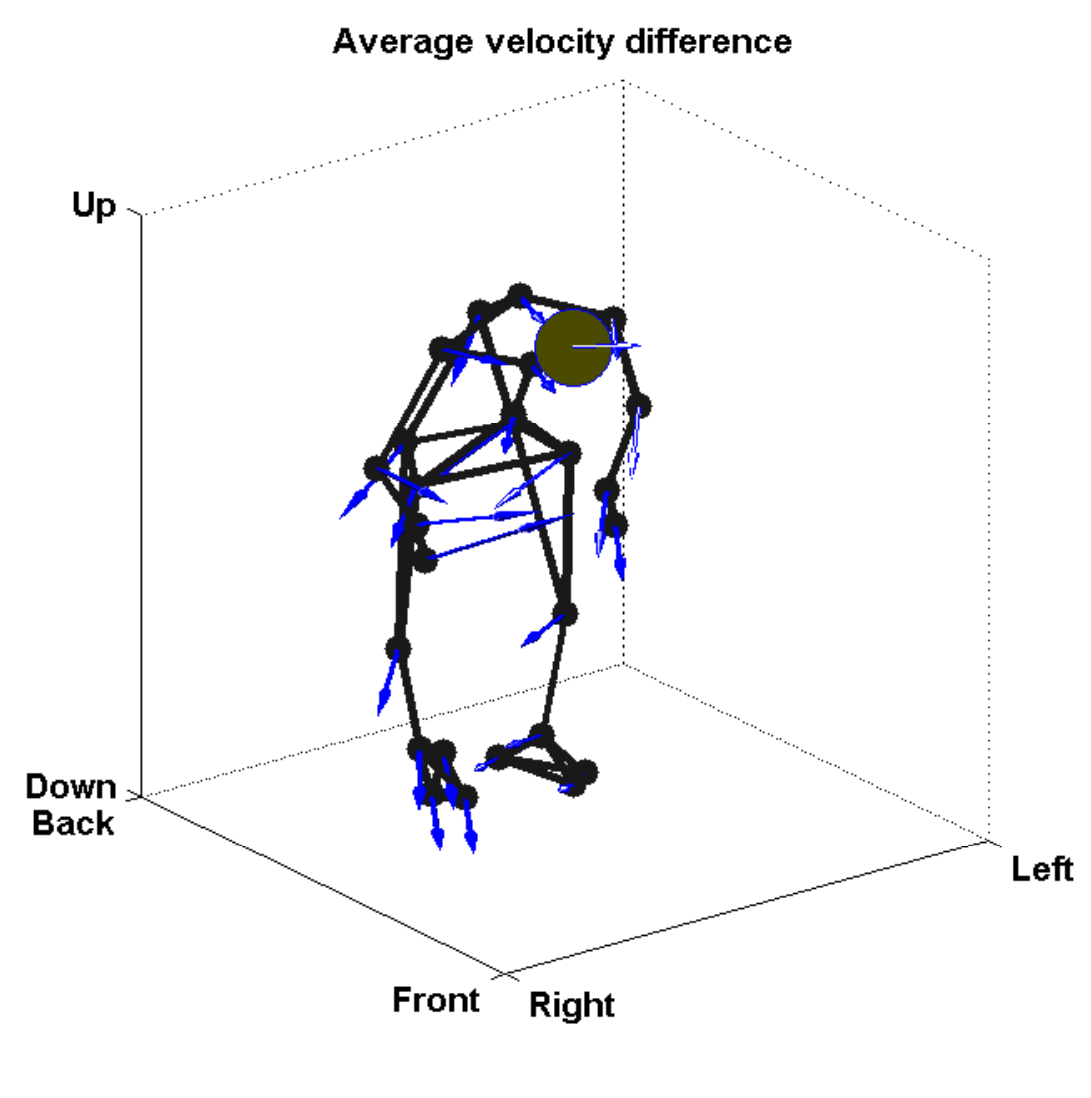}
        \caption{An illustration of the velocity changes (magnitude and direction) when converting from sadness to anger. The direction  of the arrows shows the direction of the velocity and the length of the arrows is proportional to the increase in the velocity of their corresponding joints in the generated movement. }
\label{fig:sad2ang}
\end{figure}

We have excluded the cases in which the original emotion is fear and rerun the ANOVA tests. A significant main effect of the target emotion was found for all cases at $p<0.01$. The original emotion only has a significant main effect on anger and fear ratings  after excluding movements whose original emotion is fear. 
For fear ratings, there is a large variation between ratings of fear across different original emotions (e.g., last column of Figure \ref{fig:Heat}). For the sadness to anger conversion, both anger and sadness are rated high ($p = 0.47, d = 0.035$;  sadness to anger movement in Figures \ref{fig:Heat} and \ref{fig:Sig}), which shows the main effect of the original emotion on the participants' rating of anger. 
Figure \ref{fig:sad2ang} shows the velocity changes for the sadness to anger movement used in the user study.  It shows an overall drooped body posture, which is characteristic of sadness.  Combined with the fast and sudden forward and inward movement of the hands and feet (see Figure \ref{fig:sad2ang}), we hypothesize that the mixture of typically-angry extremity movements and the drooped posture of sadness may communicate frustration, which could explain the confusion between sadness and anger for the sadness to anger movement (see the supplementary video for sadness to anger conversion).   

Nevertheless, after excluding the cases in which the original emotion is fear, in 6 out of 9 remaining cases, the target emotion is rated the highest by the participants with significant pairwise differences (at $p<0.05$) between the ratings of the target emotion and those of other emotions.  These significant pairwise differences are of medium to large size (Cohen's $d>0.7$), which indicate that the target emotion was unambiguously recognized in these cases. 
Therefore, the proposed approach is capable of generating movements with recognizable target emotions. 

In the proposed approach, the nearest neighbours to a desired motion path from a target emotion class in an available training dataset   
might not closely resemble the kinematic specificities of the desired motion path. In such cases, when $n_d = 0$, the kinematic trajectory of the generated movement might exhibit deviations from that of  the desired motion path.
To overcome this limitation, the $n_d$ parameter of the proposed approach can be tuned to force the generated movement adopt the kinematic specificities of its corresponding desired motion path.
 To illustrate the effect of encoding copies of a desired motion path in the HMM$_{NN}$ ($n_d > 0$) on the kinematic trajectory of the generated movement, we  have  generated exemplar movements by setting $n_d = 1$ and animated them side-by-side with their counterparts generated by setting $n_d = 0$ and corresponding desired motion paths (see supplementary videos). As can be seen in the supplementary videos, the generated movements more closely mimic the kinematic trajectory of their corresponding  desired motion paths when $n_d > 0$ as compared with the case where $n_d = 0$, as the HMM$_{NN}$ explicitly encode the kinematics of the desired motion path when $n_d > 0$. A document describing  the supplementary videos is also provided as a supplementary material.

The computational complexity of the nearest neighbour search in the LMA space is compared to that in the high-dimensional joint space. A  direct nearest neighbour search in the joint trajectory space is not possible due to the  variable-length movement observations. 
To this end, the nearest neighbour search in the joint trajectory space is tested based on: 1)~the Kullback–Leibler (KL) distance between movement-specific HMMs (HMM~+~KL), 2)~Euclidean distance in the space spanned by the affectively non-discriminative HMM parameters identified using the RMLR approach (HMM~+~RMLR). For the HMM~+~KL search: 1)~an HMM abstraction of the test movement is obtained: HMM~$_{test}$, 2)~a separate HMM abstraction is obtained for each one of the  training movements: HMM~$_{train_i}, i = 1,\dots,n$, where $n$ is the number of available training movements, 3)~the KL distance between the test HMM (HMM~$_{test}$) and each of the training HMMs is computed, 4)~the nearest neighbour of the test movement is  selected to be the training  movement with the smallest KL distance. For the HMM~+~RMLR search: 1)~movement-specific HMMs are obtained for the entire dataset, 2)~affectively non-discriminative HMM parameters are identified using RMLR with the elastic-net penalty, and 3)~the nearest neighbour search is performed in the space spanned by the affectively non-discriminative HMM parameters. 
The elapsed time for searching the nearest neighbours based on the LMA abstraction, HMM~+~KL, HMM~+~RMRL are 0.8$\pm$0.2 sec, 83.6$\pm$13.4 sec, and 87.7$\pm$13.6 sec, respectively. As can be seen, the proposed nearest neighbour search in the LMA space is the most computationally-efficient one.
 The efficacy of the LMA encoding for motion retrieval has been previously shown in~\cite{motion-retrieval-i3d}.

Using only the nearest neighbours of a desired motion path reduces the dimensionality of the generation problem (HMM modeling), which in turn results in reduced computational complexity of the proposed approach. In comparison with other stochastic generative approaches that are non-parametric and assume continuous latent variables (e.g., Gaussian process dynamical model \cite{Wang2008}), or multiple layers of interactive hidden states (e.g., conditional restricted Boltzmann machines \cite{Taylor:2009}), HMMs are less expensive to learn and their complexity can be tuned (number of hidden states, transition between the states, training samples, and observation distributions) to accurately model movements.  The Viterbi algorithm subsequent to the HMM modeling in the proposed generation pipeline enables identifying kinematic postures and transitions most relevant to the target emotion class and the desired motion path; hence rendering the proposed approach flexible to a wide range of kinematic and affective constraints.  

Another  favourable property of the proposed approach is that the HMM-based abstraction 
highlights specificities of the generated movements captured in hidden states (key postures) and the dynamics of transition between the states. This allows exploring how a movement is modified when converting from one emotion class to another. 
A further advantage of the proposed approach is that  generated movements can also serve as training movement exemplars to generate movements with an even larger variety of styles and encoded affective expressions. Furthermore, additional movements can be generated by a random walk in the proposed LMA space representation, and interpolating/extrapolating between the LMA representations of the movements in a labeled dataset. 

In cases where copies of a desired motion path are also encoded in the HMM~$_{NN}$, tighter joint variances can be used to limit kinematic variabilities modeled by the HMM and as a result, the generated movement will more closely mimic the spatial trajectory of the desired motion path. 
In the present work the joint variances were not controlled. 
Moreover, encoding additional copies of a desired motion path in HMM~$_{NN}$ ($n_d > 0$) enables a greater control on the spatial trajectory of the generated movement to mimic the desired motion path. 
This utility of the proposed approach  is especially  beneficial when there are no close kinematic neighbours from the target emotion class to the desired motion path in  available training datasets. 
The possibility to augment the HMM~$_{NN}$ with the desired motion path, along with the possibility to control joint variances are further favourable properties of the proposed approach and demonstrate its flexibility.
A systematic approach  for setting the $n_d$ parameter of the proposed approach 
 based on motion designer objectives is warranted  and it is a direction to be explored in the future.

The main contribution of this study is an approach for affective movement generation and while we use the UCLIC dataset that contains movement exemplars for discrete affective categories to test the proposed approach, the proposed approach can be readily applied to other types of representation of emotion. For instance, having a dataset with movement exemplars from varying levels of pleasure, arousal, and dominance dimensions, the proposed approach can be used to generate a movement with a target pleasure, arousal, and dominance level (e.g., high arousal, positive pleasure, and high dominance).

\section{Limitations}
\label{sec:limit}
The LMA Effort and Shape encoding and the subsequent RMLR identify LMA components  salient to affective expressions  and those salient to movement  kinematics. The nearest neighbour search is performed in a space spanned only by kinematically-relevant attributes. The proposed approach also offers the parameter $n_d$ that can be tuned to restrict  kinematic deviation of a modulated movement from its original kinematics. However, the present generation process lacks an approach for a systematic preservation of specific kinematic attributes of the motion.    
To address this issue, the current approach can be  extended to restrict the motion modulation process to a joint space that does not distort the kinematic characteristics of the movements that need to be  preserved. This is particularly important for functional movements where the primary task is irrelevant to the expression of affect (e.g., walking and knocking). To this end, an approach similar to the one proposed by Khatib et al.,~\cite{Khatib2004} can be adapted to limit the motion modulation to the null-space of a primary action (e.g., knocking) in the movement.
The null-space implementation of the proposed approach is particularly pertinent for application-specific robots where the primary action is irrelevant to the expression of affect and it must not be distorted in the process of converting  the robot action to display a target emotion. The proposed approach in its current implementation is best suited for purely social robots, where there is no functional task that needs to be fulfilled by the robot.

The present quantifications of the Effort components are adapted from~\cite{hachimura2005analysis, SamadaniACII2013}, in which they were shown to corroborate with  annotations provided by certified motion analysts. 
However, 
the approach used to quantify the Effort Weight of a movement based on its kinetic energy can not  accurately capture the Weight in some actions, particularly those where internal muscle activation is changing independently of movement, for example Effort actions of ``Press" and ``Dab". The Press action has Strong Weight, Direct Space, and Sustained Time, and the Dab action has Light Weight, Direct Space, and Sudden Time. The kinetic energies of the Press and Dab actions are low and high, respectively, which will lead to inaccurate Weight quantifications for these actions.
Other modalities such as facial expression and myographic activities (electromyography or mechanomyography) might be needed to measure and use the force exerted in performing a movement in order to accurately quantify the Effort Weight. Therefore, an investigation for more generic quantification methods for Effort components is warranted in the future. To this end, a quantification approach that accommodates different combinations of Effort components (i.e., Effort actions: Float, Wring, Press, Glide, Dab, Flick, Slash, Punch, and Effort drives: Vision, Spell, Passion~\cite{laban1971mastery}) is needed. 

The proposed approach requires the user to specify  an HMM configuration  (i.e., number of hidden state) and the LMA neighbourhood. Presently, an  $\epsilon$-neighbourhood  is defined as a circle centered at the desired motion path with a radius equal to 10\% of the distance from the desired motion path to the furthest movement from the target emotion class in a space spanned by the affectively non-discriminative LMA components. More hidden states enable modeling fine details of movements, which is essential for movement generation with desired expressive details. In this study, the number of hidden states were tuned by visual inspection of the quality of the generated movements. In the future, an automatic approach for tuning the number of states will be investigated. Furthermore, a systematic definition of  $\epsilon$-neighbourhood  is a direction for future investigation.

\section{Conclusions}
\label{sec:con}

An approach for automatic affective movement generation is proposed that makes use of two movement abstractions: the LMA Effort and Shape components and hidden Markov modeling.
The proposed  approach 
uses LMA nearest neighbours of a desired motion path in a labeled affective movement dataset to generate a new movement that conveys a target emotion and is kinematically close to the desired motion path. 
For a desired motion path and a target emotion,  a   systematic approach is used (regularized multinomial logistic regression) to select a subset of affectively non-discriminative LMA Effort and Shape components for the target emotion based on which the nearest neighbours for the desired motion path are found.
Therefore, the proposed approach  
does not require manual selection or definition of motion descriptors.
Furthermore, the abstract semantic representation in terms of LMA components allows for an  efficient search for nearest neighbours 
of a desired motion path as compared with the search in high-dimensional joint space. 
After identifying the nearest neighbours, the proposed approach encodes them in an HMM (HMM$_{NN})$ and uses the Viterbi algorithm to estimate the optimal state sequence for the desired motion path given the HMM of its nearest neighbours. The resulting state sequence is then used to generate a modulation of the motion path that conveys the target emotion.

The performance of the proposed approach is evaluated using a full-body affective movement dataset with movements conveying anger, sadness, happiness, and fear, and the expressivity of the generated movements are evaluated objectively using a state-of-the-art affective movement recognition model and subjectively using a perceptual user study. The generated movements were recognized at a rate of 72\% by the automatic recognition model. There are cases in the user study where the target emotions were confused with other emotions all of which either their original or target emotions is fear. Excluding these confusions, participants in the user study were able to recognize target emotions at a rate of 71\%.
Therefore, the proposed approach is capable of generating movements with recognizable target emotions.

%

\bibliographystyle{IEEEtran}
\bibliography{refs}

\begin{thebibliography}{10}
\providecommand{\url}[1]{#1}
\csname url@samestyle\endcsname
\providecommand{\newblock}{\relax}
\providecommand{\bibinfo}[2]{#2}
\providecommand{\BIBentrySTDinterwordspacing}{\spaceskip=0pt\relax}
\providecommand{\BIBentryALTinterwordstretchfactor}{4}
\providecommand{\BIBentryALTinterwordspacing}{\spaceskip=\fontdimen2\font plus
\BIBentryALTinterwordstretchfactor\fontdimen3\font minus
  \fontdimen4\font\relax}
\providecommand{\BIBforeignlanguage}[2]{{%
\expandafter\ifx\csname l@#1\endcsname\relax
\typeout{** WARNING: IEEEtran.bst: No hyphenation pattern has been}%
\typeout{** loaded for the language `#1'. Using the pattern for}%
\typeout{** the default language instead.}%
\else
\language=\csname l@#1\endcsname
\fi
#2}}
\providecommand{\BIBdecl}{\relax}
\BIBdecl

\bibitem{breazeal2004designing}
C.~Breazeal, \emph{Designing sociable robots}.\hskip 1em plus 0.5em minus
  0.4em\relax The MIT Press, 2004.

\bibitem{Pineau2003271}
\BIBentryALTinterwordspacing
J.~Pineau, M.~Montemerlo, M.~Pollack, N.~Roy, and S.~Thrun, ``Towards robotic
  assistants in nursing homes: Challenges and results,'' \emph{Robotics and
  Autonomous Systems}, vol.~42, no.~3, pp. 271 -- 281, 2003, socially
  Interactive Robots. [Online]. Available:
  \url{http://www.sciencedirect.com/science/article/pii/S0921889002003810}
\BIBentrySTDinterwordspacing

\bibitem{Hartmann2006}
B.~Hartmann, M.~Mancini, and C.~Pelachaud, ``Implementing expressive gesture
  synthesis for embodied conversational agents,'' in \emph{Gesture in
  Human-Computer Interaction and Simulation}, ser. Lecture Notes in Computer
  Science, S.~Gibet, N.~Courty, and J.-F. Kamp, Eds.\hskip 1em plus 0.5em minus
  0.4em\relax Springer Berlin Heidelberg, 2006, vol. 3881, pp. 188--199.

\bibitem{Lasseter:1987}
J.~Lasseter, ``Principles of traditional animation applied to 3d computer
  animation,'' \emph{SIGGRAPH Comput. Graph.}, vol.~21, no.~4, pp. 35--44, Aug.
  1987.

\bibitem{laban1947effort}
R.~Laban and F.~Lawrence, \emph{Effort}.\hskip 1em plus 0.5em minus 0.4em\relax
  Macdonald and Evans, 1947.

\bibitem{laban1971mastery}
R.~Laban and L.~Ullmann, \emph{The mastery of movement}.\hskip 1em plus 0.5em
  minus 0.4em\relax ERIC, 1971.

\bibitem{SamadaniACII2013}
A.~Samadani, S.~Burton, R.~Gorbet, and D.~Kuli{\'c}, ``Laban effort and shape
  analysis of affective hand and arm movements,'' in \emph{Int. Conf. ACII},
  2013, pp. 343 -- 348.

\bibitem{motion-retrieval-i3d}
M.~Kapadia, I.~Chiang, T.~Thomas, N.~I. Badler, and J.~Kider, ``{Efficient
  Motion Retrieval in Large Databases},'' in \emph{Proceedings of the symposium
  on Interactive 3D graphics and games}, ser. I3D.\hskip 1em plus 0.5em minus
  0.4em\relax ACM, 2013.

\bibitem{Kulić01072008}
D.~Kuli\'{c}, W.~Takano, and Y.~Nakamura, ``Incremental learning, clustering
  and hierarchy formation of whole body motion patterns using adaptive hidden
  {M}arkov chains,'' \emph{The International Journal of Robotics Research},
  vol.~27, no.~7, pp. 761--784, 2008.

\bibitem{SamadaniTHMS2014}
A.~Samadani, R.~Gorbet, and D.~Kulic, ``Affective movement recognition based on
  generative and discriminative stochastic dynamic models,'' \emph{IEEE
  Transactions on Human-Machine Systems}, vol.~44, no.~4, pp. 454--467, Aug
  2014.

\bibitem{Chi:2000}
D.~Chi, M.~Costa, L.~Zhao, and N.~Badler, ``The {EMOTE} model for effort and
  shape,'' in \emph{Proceedings of the 27th annual conference on Computer
  graphics and interactive techniques}, ser. SIGGRAPH '00, 2000, pp. 173--182.

\bibitem{Masuda2010}
M.~Masuda and S.~Kato, ``Motion rendering system for emotion expression of
  human form robots based on laban movement analysis,'' in \emph{IEEE RO-MAN},
  Sept 2010, pp. 324--329.

\bibitem{Nakagawa2009}
K.~Nakagawa, K.~Shinozawa, H.~Ishiguro, T.~Akimoto, and N.~Hagita, ``Motion
  modification method to control affective nuances for robots,'' in \emph{IEEE
  IROS}, 2009, pp. 5003--5008.

\bibitem{Arikan:2002}
O.~Arikan and D.~A. Forsyth, ``Interactive motion generation from examples,''
  \emph{ACM Trans. Graph.}, vol.~21, no.~3, pp. 483--490, Jul. 2002.

\bibitem{TaylorNIPS2006}
G.~W. Taylor, G.~E. Hinton, and S.~T. Roweis, ``Modeling human motion using
  binary latent variables,'' in \emph{Advances in Neural Information Processing
  Systems 19}, B.~Sch\"{o}lkopf, J.~Platt, and T.~Hoffman, Eds.\hskip 1em plus
  0.5em minus 0.4em\relax MIT Press, 2007, pp. 1345--1352.

\bibitem{unuma1995fourier}
M.~Unuma, K.~Anjyo, and R.~Takeuchi, ``Fourier principles for emotion-based
  human figure animation,'' in \emph{ACM Int. Conf. SIGGRAPH}.\hskip 1em plus
  0.5em minus 0.4em\relax ACM, 1995, pp. 91--96.

\bibitem{SamadaniSocial2013}
A.-A. Samadani, E.~Kubica, R.~Gorbet, and D.~Kuli{\'c},
  ``\BIBforeignlanguage{English}{Perception and generation of affective hand
  movements},'' \emph{\BIBforeignlanguage{English}{International Journal of
  Social Robotics}}, vol.~5, pp. 35--51, 2013.

\bibitem{Shapiro:2006}
A.~Shapiro, Y.~Cao, and P.~Faloutsos, ``Style components,'' in
  \emph{Proceedings of Graphics Interface 2006}, ser. GI ’06.\hskip 1em plus
  0.5em minus 0.4em\relax CAN: Canadian Information Processing Society, 2006,
  p. 33–39.

\bibitem{Mukai:2005}
T.~Mukai and S.~Kuriyama, ``Geostatistical motion interpolation,'' in \emph{ACM
  SIGGRAPH 2005 Papers}, ser. SIGGRAPH '05.\hskip 1em plus 0.5em minus
  0.4em\relax New York, NY, USA: ACM, 2005, pp. 1062--1070.

\bibitem{Torresani2006NIPS}
L.~Torresani, P.~Hackney, and C.~Bregler, ``Learning motion style synthesis
  from perceptual observations,'' in \emph{Advances in Neural Information
  Processing Systems 19}, B.~Sch\"{o}lkopf, J.~Platt, and T.~Hoffman,
  Eds.\hskip 1em plus 0.5em minus 0.4em\relax MIT Press, 2007, pp. 1393--1400.

\bibitem{YamazakiHSMM}
T.~{Yamazaki}, N.~{Niwase}, J.~{Yamagishi}, and T.~{Kobayashi}, ``Human walking
  motion synthesis based on multiple regression hidden semi-markov model,'' in
  \emph{2005 International Conference on Cyberworlds (CW'05)}, 2005, pp. 8
  pp.--452.

\bibitem{Tilmanne2012}
J.~Tilmanne and T.~Dutoit, ``Continuous control of style and style transitions
  through linear interpolation in hidden markov model based walk synthesis,''
  in \emph{Transactions on Computational Science XVI}, M.~L. Gavrilova and
  C.~J.~K. Tan, Eds.\hskip 1em plus 0.5em minus 0.4em\relax Berlin, Heidelberg:
  Springer Berlin Heidelberg, 2012, pp. 34--54.

\bibitem{Hsu:2005}
\BIBentryALTinterwordspacing
E.~Hsu, K.~Pulli, and J.~Popovi\'{c}, ``Style translation for human motion,''
  in \emph{ACM SIGGRAPH 2005 Papers}, ser. SIGGRAPH ’05.\hskip 1em plus 0.5em
  minus 0.4em\relax New York, NY, USA: Association for Computing Machinery,
  2005, p. 1082–1089. [Online]. Available:
  \url{https://doi.org/10.1145/1186822.1073315}
\BIBentrySTDinterwordspacing

\bibitem{Brand:2000}
M.~Brand and A.~Hertzmann, ``Style machines,'' in \emph{Proceedings of the 27th
  Annual Conference on Computer Graphics and Interactive Techniques}, ser.
  SIGGRAPH '00.\hskip 1em plus 0.5em minus 0.4em\relax ACM Press/Addison-Wesley
  Publishing Co., 2000, pp. 183--192.

\bibitem{Wang2008}
J.~Wang, D.~Fleet, and A.~Hertzmann, ``Gaussian process dynamical models for
  human motion,'' \emph{Pattern Analysis and Machine Intelligence, IEEE
  Transactions on}, vol.~30, no.~2, pp. 283--298, Feb 2008.

\bibitem{Taylor:2009}
G.~W. Taylor and G.~E. Hinton, ``Factored conditional restricted boltzmann
  machines for modeling motion style,'' in \emph{Proceedings of the 26th Annual
  International Conference on Machine Learning}, ser. ICML '09.\hskip 1em plus
  0.5em minus 0.4em\relax New York, NY, USA: ACM, 2009, pp. 1025--1032.

\bibitem{Chiu:2011}
C.-C. Chiu and S.~Marsella, ``A style controller for generating virtual human
  behaviors,'' in \emph{The 10th International Conference on Autonomous Agents
  and Multiagent Systems - Volume 3}, ser. AAMAS ’11.\hskip 1em plus 0.5em
  minus 0.4em\relax Richland, SC: International Foundation for Autonomous
  Agents and Multiagent Systems, 2011, p. 1023–1030.

\bibitem{HintonRBMGuide}
G.~Hinton, ``A practical guide to training restricted boltzmann machines,'' in
  \emph{Neural Networks: Tricks of the Trade}, ser. Lecture Notes in Computer
  Science, G.~Montavon, G.~Orr, and K.-R. Müller, Eds.\hskip 1em plus 0.5em
  minus 0.4em\relax Springer Berlin Heidelberg, 2012, vol. 7700, pp. 599--619.

\bibitem{bartenieff1965effort}
I.~Bartenieff, \emph{Effort-Shape analysis of movement: The unity of expression
  and function}.\hskip 1em plus 0.5em minus 0.4em\relax Albert Einstein College
  of Medicine, Yeshiva University, 1965.

\bibitem{hackney2004making}
P.~Hackney, \emph{Making connections: Total body integration through Bartenieff
  fundamentals}.\hskip 1em plus 0.5em minus 0.4em\relax Routledge, 2004.

\bibitem{hachimura2005analysis}
K.~Hachimura, K.~Takashina, and M.~Yoshimura, ``Analysis and evaluation of
  dancing movement based on {LMA},'' in \emph{IEEE International Workshop on
  Robot and Human Interactive Communication, ROMAN.}\hskip 1em plus 0.5em minus
  0.4em\relax IEEE, 2005, pp. 294--299.

\bibitem{dell1977primer}
C.~Dell, \emph{A primer for movement description using effort-shape and
  supplementary concepts}.\hskip 1em plus 0.5em minus 0.4em\relax Princeton
  Book Company Pub, 1977.

\bibitem{nakata2002analysis}
T.~Nakata, T.~Mori, and T.~Sato, ``Analysis of impression of robot bodily
  expression,'' \emph{Journal of Robotics and Mechatronics}, vol.~14, no.~1,
  pp. 27--36, 2002.

\bibitem{lamb1965posture}
W.~Lamb, \emph{Posture and gesture: An introduction to the study of physical
  behaviour}.\hskip 1em plus 0.5em minus 0.4em\relax G. Duckworth, 1965.

\bibitem{Friedman2010}
J.~H. Friedman, T.~Hastie, and R.~Tibshirani, ``Regularization paths for
  generalized linear models via coordinate descent,'' \emph{Journal of
  Statistical Software}, vol.~33, no.~1, pp. 1--22, 2 2010.

\bibitem{hastie2009elements}
T.~Hastie, R.~Tibshirani, and Friedman, \emph{The elements of statistical
  learning}, 2nd~ed., ser. Springer series in statistics.\hskip 1em plus 0.5em
  minus 0.4em\relax Springer, 2009.

\bibitem{RabinerHMM}
L.~Rabiner, ``A tutorial on hidden {M}arkov models and selected applications in
  speech recognition,'' \emph{Proceedings of the IEEE}, vol.~77, no.~2, pp. 257
  --286, feb 1989.

\bibitem{Beyer1999}
K.~Beyer, J.~Goldstein, R.~Ramakrishnan, and U.~Shaft, ``When is ``nearest
  neighbor'' meaningful?'' in \emph{Database Theory --- ICDT'99}, C.~Beeri and
  P.~Buneman, Eds.\hskip 1em plus 0.5em minus 0.4em\relax Berlin, Heidelberg:
  Springer Berlin Heidelberg, 1999, pp. 217--235.

\bibitem{kleinsmith2006cross}
A.~Kleinsmith, P.~De~Silva, and N.~Bianchi-Berthouze, ``Cross-cultural
  differences in recognizing affect from body posture,'' \emph{Interacting with
  Computers}, vol.~18, no.~6, pp. 1371--1389, 2006.

\bibitem{friedmanglmnet}
J.~Qian, T.~Hastie, J.~Friedman, R.~Tibshirani, and N.~Simon, ``Glmnet for
  matlab 2010,'' \url{http://www.stanford.edu/~hastie/glmnet_matlab/}.

\bibitem{pkekalska2005dissimilarity}
E.~Pekalska and R.~Duin, \emph{The dissimilarity representation for pattern
  recognition: foundations and applications}.\hskip 1em plus 0.5em minus
  0.4em\relax World Scientific Publishing Company, 2005, vol.~64.

\bibitem{faul2007g}
F.~Faul, E.~Erdfelder, A.~Lang, and A.~Buchner, ``G* power 3: A flexible
  statistical power analysis program for the social, behavioral, and biomedical
  sciences,'' \emph{Behavior research methods}, vol.~39, no.~2, pp. 175--191,
  2007.

\bibitem{cohen2013statistical}
\BIBentryALTinterwordspacing
J.~Cohen, \emph{Statistical Power Analysis for the Behavioral Sciences}.\hskip
  1em plus 0.5em minus 0.4em\relax Taylor \& Francis, 2013. [Online].
  Available: \url{http://books.google.ca/books?id=cIJH0lR33bgC}
\BIBentrySTDinterwordspacing

\bibitem{Russell1980}
J.~Russell, ``A circumplex model of affect,'' \emph{PERS SOC PSYCHOL}, vol.~39,
  pp. 1161 -- 1178, 1980.

\bibitem{van2007body}
J.~Van~den Stock, R.~Righart, and B.~de~Gelder, ``Body expressions influence
  recognition of emotions in the face and voice.'' \emph{Emotion}, vol.~7,
  no.~3, pp. 487--494, 2007.

\bibitem{montepare1987}
J.~Montepare, S.~Goldstein, and A.~Clausen, ``The identification of emotions
  from gait information,'' \emph{J NONVERBAL BEHAV}, vol.~11, pp. 33--42, 1987.

\bibitem{+Coulson2004}
M.~Coulson, ``Attributing emotion to static body postures: Recognition
  accuracy, confusions, and viewpoint dependence,'' \emph{Nonverbal Behavior},
  vol.~28, no.~2, pp. 117--139, 2004.

\bibitem{Khatib2004}
\BIBentryALTinterwordspacing
O.~KHATIB, L.~SENTIS, J.~PARK, and J.~WARREN, ``Whole-body dynamic behavior and
  control of human-like robots,'' \emph{International Journal of Humanoid
  Robotics}, vol.~01, no.~01, pp. 29--43, 2004. [Online]. Available:
  \url{https://doi.org/10.1142/S0219843604000058}
\BIBentrySTDinterwordspacing

\end{thebibliography}

\end{document}